\documentclass[reprint,superscriptaddress,amsmath,amssymb,aps,prl,floatfix,longbibliography]{revtex4-2}

\usepackage{hyperref}
\usepackage{nameref}

\usepackage{graphicx}
\graphicspath{{figures/}}
\usepackage{amsmath}
\usepackage{amssymb}
\usepackage{multirow} 
\usepackage{physics} 

\usepackage{mathtools}

\usepackage{braket}
\usepackage{bbold}
\usepackage{xcolor}
\usepackage{booktabs}
\usepackage[normalem]{ulem}

\usepackage[section]{placeins}
\providecommand{\ignore}[1]{}

\newif\ifcmnt
\cmnttrue
\ifdefined\cmntsoff\cmntfalse\fi

\ifcmnt
    \providecommand{\aucmnt}[1]{#1}

\else
    \providecommand{\aucmnt}[1]{}

\fi

\graphicspath{{figures/}}

\AtBeginDocument{\let\oldcontentsline\contentsline}
\newcommand{\notoccontentsline}[4]{\oldcontentsline{}{}{}{}}
\newcommand{\droptocpage}{\addtocontents{toc}{\let\protect\contentsline\protect\notoccontentsline}}
\newcommand{\incltocpage}{\addtocontents{toc}{\let\protect\contentsline\protect\oldcontentsline}}

\begin{document}
\title{An atomic boson sampler}







\author{Aaron W. Young}
\affiliation{JILA, University of Colorado and National Institute of Standards and Technology, and Department of Physics, University of Colorado, Boulder, CO 80309, USA}

\author{Shawn Geller}
\affiliation{National Institute of Standards and Technology, Boulder, CO 80305, USA}
\affiliation{Department of Physics, University of Colorado, Boulder, CO 80309, USA}

\author{William J. Eckner}
\affiliation{JILA, University of Colorado and National Institute of Standards and Technology, and Department of Physics, University of Colorado, Boulder, CO 80309, USA}

\author{Nathan Schine}
\affiliation{JILA, University of Colorado and National Institute of Standards and Technology, and Department of Physics, University of Colorado, Boulder, CO 80309, USA}
\affiliation{Joint Quantum Institute, University of Maryland Department of Physics and National Institute of Standards and Technology, College Park, MD 20742}

\author{Scott Glancy}
\affiliation{National Institute of Standards and Technology, Boulder, CO 80305, USA}

\author{Emanuel Knill}
\affiliation{National Institute of Standards and Technology, Boulder, CO 80305, USA}
\affiliation{Center for Theory of Quantum Matter, University of Colorado, Boulder, CO 80309, USA.}

\author{Adam M. Kaufman}
\affiliation{JILA, University of Colorado and National Institute of Standards and Technology, and Department of Physics, University of Colorado, Boulder, CO 80309, USA}
\email[E-mail:]{adam.kaufman@colorado.edu}

\date{\today}

\maketitle

\begin{figure}
	\includegraphics[width=\linewidth]{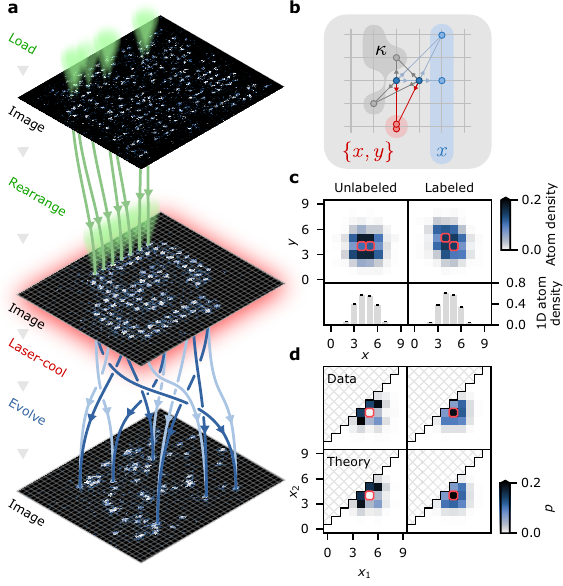}
  \caption{\textbf{Assembling Fock states of bosonic atoms in a tunnel-coupled optical lattice.} \textbf{a}, 
  States containing up to 180 bosonic atoms are prepared and measured on-demand in an optical lattice (grey grid) via site- and atom-resolved imaging (pictured images are single-shot experimental data), parallelized rearrangement with optical tweezers (green cones), and high fidelity laser cooling.
	The evolution of these atoms in the lattice is described by factorially many (in particle number) different interfering multiparticle trajectories (light and dark blue lines show two possible interfering trajectories for a subset of atoms).
	\textbf{b}, The interference between atoms can be studied by measuring correlations in the occupation of specific sites in the lattice (red).
  Partitioning the lattice into arbitrary subsets of sites $\kappa$ (grey) can allow one to more efficiently (in terms of the number of experimental trials) study the effects of interference.
  Due to the separability of quantum walk dynamics on an ideal square lattice, it is particularly useful to partition the lattice into columns indexed by their $x$ coordinates (blue).
  \textbf{c}, Specifically, by binning 2D quantum walk data over the $y$ axis, one can study 1D quantum walks along $x$, where the initial coordinate in the $y$ axis acts as an additional ``hidden'' degree of freedom that can modify the distinguishability of the atoms in the remaining ``visible'' $x$ axis.
  For example, one can prepare two atoms with the same (unlabeled) or different (labeled) $y$ coordinates (sites initialized with an atom are marked in red).
  The atom number density on each site after the evolution of these atoms is shown here at an evolution time of $t = t_{\rm HOM} \coloneqq 0.96$~ms.
	\textbf{d}, After binning to 1D, each two-particle output is uniquely labeled by the $x$ coordinates of the two atoms $(x_1,x_2)$, with $x_2\le x_1$.
  The measured probabilities $p$ of observing a given two-particle output are in agreement with theory, where the unlabeled atoms (top left) agree with the expectation for perfectly indistinguishable bosons (bottom left), and labeled atoms (top right) with distinguishable particles (bottom right).
  In these and subsequent probability distributions in this work, the input state before the atom evolution is highlighted in red.
  }
\label{fig:1}
\end{figure}

A boson sampler implements a restricted model of quantum computing.
It is defined by the ability to sample from the distribution resulting from the interference of identical bosons propagating according to programmable, non-interacting dynamics~\cite{aaronson_computational_2011}.
An efficient exact classical simulation of boson sampling is not believed to exist, which has motivated ground-breaking boson sampling experiments in photonics with increasingly many photons~\cite{peruzzo_quantum_2010, sansoni_two-particle_2012,
broome_photonic_2013, spring_boson_2013, tillmann_experimental_2013, bentivegna_experimental_2015, carolan_universal_2015, wang_boson_2019, zhong_quantum_2020, madsen_quantum_2022, deng_gaussian_2023}.
However, it is difficult to generate and reliably evolve specific numbers of photons with low loss, and thus probabilistic techniques for postselection~\cite{bentivegna_experimental_2015} or significant changes to standard boson sampling~\cite{zhong_quantum_2020, madsen_quantum_2022, deng_gaussian_2023} are generally employed.
Here, we address the above challenges by implementing boson sampling using ultracold atoms~\cite{muraleedharan_quantum_2019, robens_boson_2022} in a two-dimensional, tunnel-coupled optical lattice.
This demonstration is enabled by a new combination of tools involving high fidelity optical cooling and imaging of atoms in a lattice, as well as programmable control of those atoms using optical tweezers.
When extended to interacting systems, our work demonstrates the core capabilities required to directly assemble ground and excited states in simulations of various Hubbard models~\cite{lee_doping_2006, bloch_many-body_2008}.

There is a rich history of studying quantum optics with atoms instead of photons~\cite{cronin_optics_2009}, including demonstrations of two-atom interference~\cite{kaufman_two-particle_2014, lopes_atomic_2015,islam_measuring_2015, robens_boson_2022}.
Scaling these demonstrations up requires reliable preparation of chosen patterns of many identical atoms, evolution under non-interacting dynamics that can exchange the positions of the atoms, and high-fidelity detection of the atom positions after their evolution.
The latter two requirements can be met with quantum gas microscopy, where it is possible to prepare and detect thousands of atoms in an optical lattice in which those atoms can tunnel and interfere~\cite{bakr_quantum_2009, sherson_single-atom-resolved_2010}.
While in principle the first requirement can be met by additionally using sophisticated optical control techniques to isolate a subset of atoms from a larger many-body state~\cite{weitenberg_single-spin_2011, murmann_two_2015, islam_measuring_2015, kaufman_quantum_2016, yan_two-dimensional_2022-1, zheng_efficiently_2022}, in practice, the typical state fidelities and cycle times of up to tens of seconds make it challenging to perform a large-scale boson sampling demonstration.
To explore alternative routes to improve the speed and quality of state preparation, we turn to tools that have been developed to rapidly image~\cite{covey_2000-times_2019}, optically cool~\cite{kaufman_cooling_2012, thompson_coherence_2013, norcia_microscopic_2018, cooper_alkaline-earth_2018}, and deterministically rearrange individual atoms trapped in optical tweezers~\cite{endres_atom-by-atom_2016, barredo_atom-by-atom_2016}.
Using the tweezers to implant atoms in a tunnel-coupled optical lattice~\cite{trisnadi_design_2022, young_tweezer-programmable_2022} allows for both fast state preparation (in hundreds of milliseconds), and the required dynamics for boson sampling.
Moreover, the tweezers can be used to modify the lattice potential~\cite{young_tweezer-programmable_2022} to implement different non-interacting dynamics with low loss.

Combining the aforementioned tools for state preparation, evolution, and detection in a single apparatus enables large-scale demonstrations of boson sampling with Fock states that were not feasible prior to this work.
In particular, we study specific instances of boson sampling involving up to 180 atoms spread over $\sim 1015$ sites in the lattice.
Importantly, direct verification of boson sampling is infeasible for these system sizes.
In order to build confidence that these boson sampling experiments do indeed sample from a very large state space and are not feasible to simulate classically, we develop and implement broadly applicable tests to examine and quantify the performance of our system for boson sampling.
These tests include (i) stringent tests of indistinguishability with up to eight atoms; (ii) characterizations of the atom evolution from data with one and two atoms; (iii) evidence of expected bunching features as a result of interference for a range of atom numbers and effective particle statistics.
Critically, this diversity of tests is enabled by the programmability of the input states available in our system, the high repetition rate of $\sim1$~Hz, and the family of different unitaries that can be realized using Hamiltonian evolution on the lattice.

We begin with a description of our experimental setup and its features as they relate to the tests and demonstrations performed (Fig.~\ref{fig:1}).
To prepare patterns of identical $\rm ^{88}Sr$ atoms in an optical lattice, we load a thermal cloud of atoms into a tweezer array (typically with dimensions of $16\times24$) with random filling of 50-75~\%~\cite{young_half-minute-scale_2020, jenkins_ytterbium_2022-1}, and implant these atoms in the lattice~\cite{schine_long-lived_2022, young_tweezer-programmable_2022}.
We image the implanted atoms with a typical fidelity of $99.8(1)\;\%$ per lattice site, and, based on these images, rearrange the atoms into the desired target patterns (Fig.~\ref{fig:1}a).
The rearrangement can be performed with a per-atom success probability as high as $99.5\;\%$, but $98\;\%$ is typical for the data presented in this work~(see Methods).
A second image is used to check if the atoms have been rearranged correctly, allowing us to postselect for perfect atom rearrangement (up to imaging errors).
After the atoms are appropriately positioned in the lattice, they are cooled to their three-dimensional (3D) motional ground state via resolved sideband cooling using the narrow line $\rm {^1S_0} \leftrightarrow {^3P_1}$ transition (see Methods)~\cite{young_tweezer-programmable_2022}.
Note that similar performance for state preparation and imaging is achieved across a $48\times48$-site region in the lattice (see Supplementary Materials [SM] Sec.~\ref{sup:calibration}).

Once prepared, each atom undergoes quantum walk dynamics described by an $m\times m$ single particle unitary $U = e^{-i H t/\hbar}$, where $m$ is the number of sites in the lattice, $H$ the lattice Hamiltonian (see Methods), and $t$ the evolution time.
$5.0(2)\;\%$ of the atoms are lost during the quantum walk dynamics, which we refer to as single-particle loss (see Methods).
Critically, single particle loss is not strongly dependent on evolution time (see SM Sec.~\ref{sup:calibration}), and is primarily due to state preparation errors where atoms occupying excited in-plane bands leave the analysis region and are unlikely to return.
After the quantum walk, a final image is used to measure the positions of the atoms.
This measurement is not number resolving, and instead detects atom number parity on each site due to the effect of light assisted collisions (we refer to this process as parity projection)~\cite{schlosser_sub-poissonian_2001, bakr_quantum_2009,sherson_single-atom-resolved_2010}.
However, for most experiments in this work we can operate in a regime where the probability that more than one atom occupies a given site is low, minimizing the effect of parity projection~\cite{aaronson_computational_2011}.
Additionally, except for experiments that involve large atom numbers, we postselect our measurements on observing no lost or extra atoms after the quantum walk dynamics to account for any lingering effects of loss, image infidelity, and parity projection (see Methods).

Our use of ground state $\rm ^{88}Sr$ atoms ensures that on-site elastic and inelastic interactions are weak~(see SM Sec.~\ref{sup:interactions})~\cite{martinez_de_escobar_two-photon_2008,goban_emergence_2018}, and that the many atom dynamics are, to a good approximation, non-interacting.
For $n$ non-interacting bosonic atoms, the many-atom evolution is related to the permanent of an $n\times n$ submatrix of $U$ (see Methods).
Computing the permanent of an arbitrary matrix is in the complexity class \#P-hard~\cite{aaronson_computational_2011,valiant_complexity_1979}, and so even sampling from the probability distribution resulting from the above evolution is believed to be intractable for more than ${\sim 50}$ atoms using classical techniques~\cite{clifford_classical_2018}.
By contrast, for distinguishable atoms the evolution is related to the permanent of an $n\times n$ submatrix of $|U|^2$, where $|\cdot|$ denotes the element-wise norm (see Methods).
Approximating the permanent of a non-negative real matrix can be performed in polynomial time, and the corresponding sampling task can efficiently be accomplished classically~\cite{tichy_sampling_2015}.
Different degrees of atom distinguishability result in behaviors that lie in between these two scenarios~\cite{tichy_sampling_2015, dufour_many-body_2020}.

Although our atoms are fundamentally bosonic composite particles, they may not behave bosonically on the lattice.
This is because the single-atom state space includes degrees of freedom (DOFs) other than just the location of the atom in the lattice.
For example, the atom can be in different electronic states, or in different motional states in the direction that is normal to the lattice.
The single atom state space is thus a tensor product of the Hilbert spaces $\mathcal{H}_{V}$ (``visible'' DOFs) spanned by $\ket{i}$ (sites), and $\mathcal{H}_{H}$ (``hidden'' DOFs) spanned by $\ket{h}$ (labels).
In our system, the overall unitary dynamics is non-interacting for the atoms, and the single atom Hamiltonian acts independently on the hidden and the visible DOFs, which means that it is of the form $H_{V}\otimes \mathbb{1}_H + \mathbb{1}_V \otimes H_{H}$ (where the subscripts indicate the subspace that each operator acts on, and $H$ is a Hamiltonian acting on the indicated subspace).
Due to this independence, the hidden DOFs can affect the visible behavior of the atoms only by changing their effective particle statistics.
For example, if we prepare some number of atoms in specific visible sites but each atom has a different hidden label, then the visible behavior of the atoms is that of perfectly distinguishable particles.
Other types of visible particle statistics are also possible, including fermionic statistics if the multi-particle wavefunction is antisymmetric in the hidden DOFs~\cite{sansoni_two-particle_2012}.
Our tests focus on determining limits on the deviation from bosonic behavior due to the hidden DOFs.
We also characterize the specific particle statistics exhibited in the visible behavior in experiments with only two and three particles.
Note that errors in the visible DOFs are unlikely due to our procedure for state preparation and postselection~(see Methods), which rules out certain exotic error models~\cite{tichy_stringent_2014}.

\begin{figure*}
	\includegraphics[width=\linewidth]{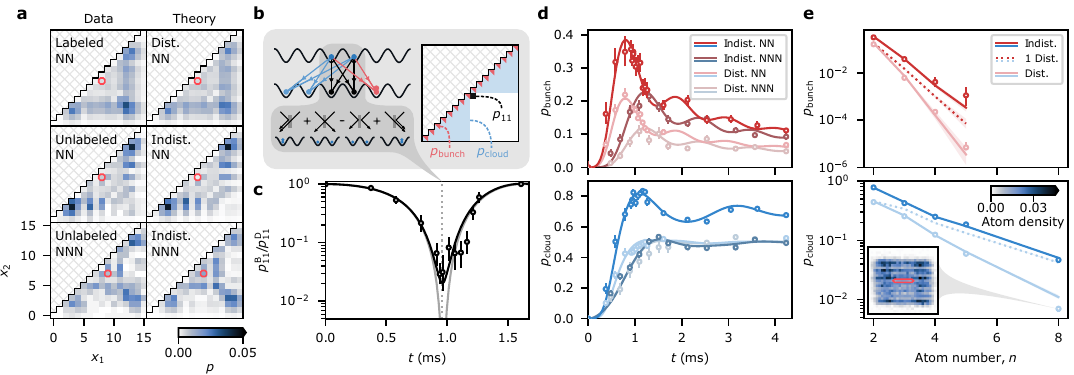}
  \caption{\textbf{Multiparticle quantum walks in 1D.} 
  \textbf{a}, Quantum walks in 1D, pictured here for two particles at an evolution time of $t = 4.23$~ms for different input states and particle statistics.
  The atoms are prepared in either neighboring sites in the lattice (nearest-neighbor, NN), or separated by one site (next-nearest-neighbor, NNN).
  The theoretical predictions are for perfectly distinguishable particles (Dist.) and indistinguishable bosons (Indist.).
  The associated measurements either introduce (Labeled) or do not introduce (Unlabeled) a time label to distinguish between the different atoms.
  In subsequent figures we use ``Dist.'' or ``Indist.'' to refer to both data and theory for consistency, with the understanding that the experimental measurements may not correspond to perfectly indistinguishable bosons.
  \textbf{b}, We consider three ways of coarse-graining multi-atom distributions: $p_{11}$ refers to the probability of coincident detection of one atom on each of two input sites (black output), $p_\mathrm{bunch}$ refers to a coincident detection of all atoms on the same site (summing over red outputs), and $p_\mathrm{cloud}$ refers to all atoms appearing on the same half of the array (summing over blue outputs).
  Example trajectories that fulfill these conditions are shown on the left. 
  \textbf{c}, We plot the ratio $p_{11}^{\rm B}/p_{11}^{\rm D}$, where the superscripts $D$ and $B$ respectively refer to experiments with and without introducing an additional label for distinguishability.
  At an evolution time of $t=t_{\rm HOM}$ (grey dotted line, callout in \textbf{b}) this is analogous to measurements of Hong-Ou-Mandel (HOM) interference using a balanced beam splitter.
  Note that although the HOM dip can have unity visibility for identical bosons in an ideal lattice with only nearest-neighbor tunneling (grey theory curve), higher order tunneling terms in our lattice~\cite{young_tweezer-programmable_2022} result in imperfect visibility even for identical bosons (black theory curve).
  \textbf{d}, Measurements of $p_\mathrm{bunch}$ and $p_\mathrm{cloud}$ for two atoms as a function of evolution time, and for both NN and NNN inputs, are in good agreement with theory. 
  \textbf{e}, Measurements of $p_\mathrm{bunch}$ and $p_\mathrm{cloud}$ for NN input patterns (with up to $n = 5$ and 8 atoms respectively) are also in good agreement with theory.
  For comparison, we display a prediction for when each trial of the experiment contains one randomly selected atom that is distinguishable from the rest (1 Dist.).
  The inset shows the measured atom density in 2D with 8 prepared atoms, illustrating how at the chosen evolution time of $t = (n-1) t_{\rm HOM}$ for the measurements in \textbf{e}, all $n$ input sites are approximately uniformly coupled to each other.
  The theory predictions appearing throughout this figure are for error-free preparations of atoms with the indicated particle statistics.
  The theory curves in \textbf{e} are performed using Monte-Carlo methods, and thus the shaded regions denote $\pm1\sigma$ confidence intervals that include both sampling errors, and systematic errors relating to fluctuations in the applied unitary evolution~(see SM Sec.~\ref{sup:error_model}).
  } \label{fig:2}
\end{figure*}

The indistinguishability of two atoms is defined to be the Hilbert-Schmidt inner product of the density matrices of the two one-atom wavefunctions on the hidden DOFs. 
It can be inferred from a Hong-Ou-Mandel (HOM) experiment by comparing the coincidence probability of the atoms $p_{11}^{\rm B}$ to the corresponding coincidence probability for perfectly distinguishable atoms $p_{11}^{\rm D}$ (see Methods).
We measure HOM interference of atoms~\cite{kaufman_two-particle_2014, lopes_atomic_2015, robens_boson_2022} by studying quantum walks in 1D via the binning procedure described in the Methods (Fig.~\ref{fig:2}).
In these quantum walks, an evolution time of $t_{\rm HOM} \coloneqq 0.96$~ms approximates a balanced beam splitter between adjacent sites (see Fig.~\ref{fig:2}bc and Methods).
To measure the behavior of distinguishable atoms, we can control the hidden (after binning) initial $y$ coordinates of the atoms via rearrangement (we refer to this as position labeling. See Fig.~\ref{fig:1}cd).
Alternatively, we can perform a pair of experiments where only one atom is prepared at a time, and then combine the data in subsequent analysis (we refer to this as time labeling. See Methods).
Both approaches agree to within statistical errors.

Using the above techniques, we measure the contrast of the HOM dip to be $97.1^{+1.0}_{-1.5}\;\%$, which serves as a lower bound on the atom indistinguishability.
We estimate the indistinguishability to be $99.5^{+0.5}_{-1.6}\;\%$ given additional modelling of our lattice potential (see Methods and SM Sec.~\ref{sup:calibration}).
We suspect that the dominant source of distinguishability is imperfect cooling in the direction normal to the lattice~(see Methods and SM Sec.~\ref{sup:budget}).
We also performed a separate measurement of the indistinguishability that does not rely on either postselection or binning, and found that the two measurements were consistent (see SM Sec.~\ref{sup:loss2indist}).
Our HOM measurements are averaged over three different regions in the lattice, with similar performance attained across a region that contains all input sites used in this work~(see SM secs.~\ref{sup:calibration} and \ref{sup:stats}).

\begin{figure}
	\includegraphics[width=\linewidth]{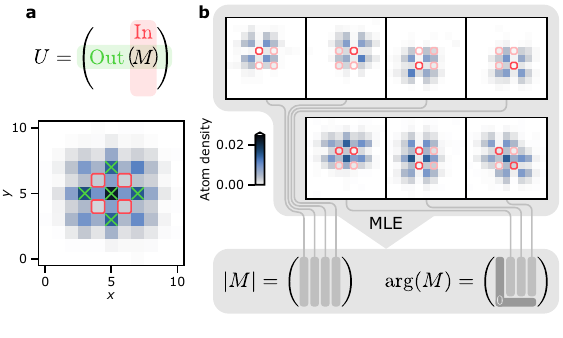}
  \caption{\textbf{Characterizing the single particle unitary.} \textbf{a}, 
  We directly characterize the evolution of atoms from the input sites circled in red to the output sites marked by the green crosses. This evolution is determined by a $5 \times 4$ submatrix $M$ of the single particle unitary $U$.
  \textbf{b}, A series of different one- and two-particle quantum walks are used to infer $M$. Density plots associated with measurements of these different quantum walks are shown in the upper panel, with input sites that are populated in a given preparation circled in dark red (unpopulated input sites in light red).
  $M$ is estimated using the maximum likelihood estimation (MLE) procedure described in the text.
  Schematically, different preparations provide information about different elements in $M$, as indicated by the grey lines~\cite{laing_super-stable_2012}.
  The color scale is shared across all parts of this figure.
  }
\label{fig:3}
\end{figure}


One way to test the visible particle statistics of more than two atoms is to measure full bunching~\cite{spagnolo_general_2013}, which is the probability \(p_{\mathrm{bunch}}\) that all atoms occupy the same visible output site. 
If the visible and hidden dynamics are perfectly separable, full bunching is uniquely maximized by bosonic particle statistics.
Note that in order to account for the effect of parity projection, we normalize the bunching probability of distinguishable particles by the probability of full survival for bosonic particles (see SM Sec.~\ref{sup:distinguishableconfint_bunchcloud}).  
The results of our full bunching measurements are shown in Fig.~\ref{fig:2}de, and are in good agreement with theoretical predictions.
Although the lattice Hamiltonian in our experiments is not perfectly separable, this agreement provides strong evidence for the indistinguishability of the atoms.

The full bunching probability decreases very rapidly with atom number, and is infeasible for us to measure for more than 5 atoms.
In order to test the indistinguishability of larger numbers of atoms, it is beneficial to look for signals that are sensitive to interference, but converge more quickly.
Two quantities of interest are clouding and modified generalized bunching, defined in the Methods.
Both of these quantities can be thought of as a way to quantify a general tendency for the atoms to end up on the same site, or (in the case of clouding) close together.
Unlike full bunching, interpreting measurements of clouding and modified generalized bunching requires precise knowledge of the atom evolution, namely of the single-particle unitary $U$.

For all theory predictions appearing in this work, we perform spectroscopic measurements that use the atoms as local probes of the lattice depth to generate a model of the lattice Hamiltonian and thus of $U$~(see SM Sec.~\ref{sup:calibration}).
However, more direct measurements of $U$ are important for future studies that attempt to program $U$ using local, possibly time-varying potentials imposed by the optical tweezers~\cite{muraleedharan_quantum_2019, young_tweezer-programmable_2022}. 
As a proof of principle, we show that $U$ can be inferred directly by measuring the interference resulting from different preparations of Fock states~\cite{laing_super-stable_2012} (Fig.~\ref{fig:3}).
Specifically, we perform a maximum likelihood (ML) fit of the parameters of interest in the unitary using data involving preparations of one and two atoms in the lattice, which results in better precision than previous approaches (see Methods)~\cite{laing_super-stable_2012}.
Using this fitting procedure, we perform characterizations of the terms in $U$ corresponding to four input sites and five output sites at an evolution time of $t = 1.46$~ms (Fig.~\ref{fig:3}).
We found that the ML fit is within statistical variation of the spectroscopic characterization of $U$ (see Extended Data Fig.~\ref{efig:2}e).
In principle, these fits allow one to characterize all parameters in $U$ at constant precision with a number of experimental trials that scales polynomially in $m$.
However, the number of parameters in $U$ that we can accurately infer using quantum walk data is currently limited by the cycle time of our experiment and drifts in the experiment that modify $U$ on long timescales.

Given the above calibrations of $U$, we can compare measurements of clouding with up to $n = 8$ atoms at an evolution time of $t = (n-1) t_{\rm HOM} $ to theory (Fig.~\ref{fig:2}de).
The behavior of the atoms is in line with the prediction for ideal bosons and is clearly separated from both measurements with time-labeled (and thus distinguishable) atoms and from theoretical predictions for partially distinguishable atoms.

\begin{figure}
	\includegraphics[width=\linewidth]{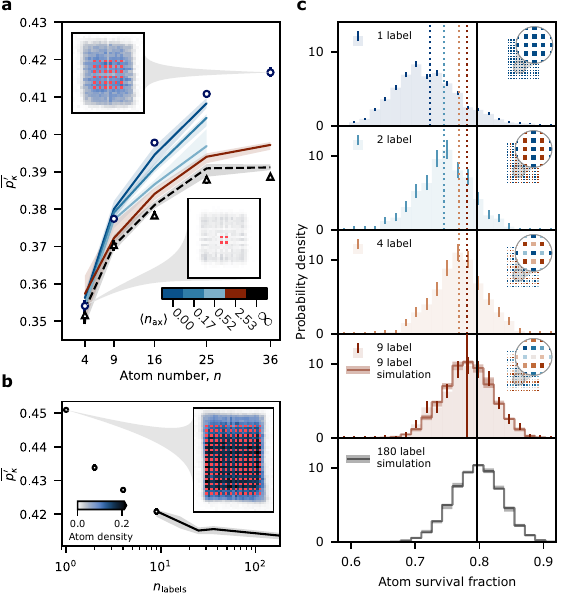}
  \caption{\textbf{Interference of large bosonic Fock states.} \textbf{a}, Measurements of the modified generalized bunching probability $\overline{p_{\kappa}'}$ as a function of atom number for square $\sqrt{n}\times \sqrt{n}$ input patterns at next-nearest-neighbor spacing and at a fixed evolution time of $t = 6.45$~ms with (triangles) and without (circles) introducing time labels that make the atoms fully distinguishable.
  These measurements can be compared to a model for partial distinguishability (colored lines) which is described by a harmonic oscillator with an expected thermal occupation of $\langle n_{\rm ax} \rangle$~(see Methods).
  The fully distinguishable case (dashed line) corresponds to infinite temperature.
	Note that simulations for low temperatures and large atom numbers are absent due to computational overhead.
  Insets in \textbf{a} and \textbf{b} show the atom density after their evolution, and share a color bar.
	\textbf{b}, Measurements of $\overline{p_{\kappa}'}$ with input patterns containing 180 atoms as a function of distinguishability (circles).
  The distinguishability is controlled by partitioning the input state into $n_\mathrm{labels}$ sub-ensembles, such that only atoms within a sub-ensemble share a time label and thus can interfere.
  For particles that are sufficiently distinguishable to simulate, we can compare these measurements to theory (solid line).
	\textbf{c}, The distribution in the observed fraction of surviving atoms on each shot of the experiment is also sensitive to the effects of interference due to parity projection.
	For the case of 9 labels (second from bottom panel), simulations capture the measured distribution of atom survival probabilities.
  Measurements with different numbers of labels (and thus atom distinguishability) are clearly resolved both from each other, and from a simulation of the fully distinguishable case with 180 labels (bottom panel).
  The vertical lines denote the mean of each distribution (dotted lines are measurements, solid lines are theory).
	Insets in \textbf{c} denote the relevant assignment of labels for each data set in both \textbf{b} and \textbf{c}, with each color corresponding to a unique time label for a subset of input sites.
  Shaded regions about all theory curves in this figure denote $\pm1\sigma$ confidence intervals, including systematic errors relating to fluctuations in the applied unitary.
  }\label{fig:4}
\end{figure}

To go beyond 8 atoms, we do not perform binning, and instead study the modified generalized bunching probability $\overline{p_{\kappa}'}$ (see Methods) as a function of atom number for square $\sqrt{n}\times\sqrt{n}$ input patterns with next-nearest neighbor spacing, and at a fixed evolution time of $t = 6.45$~ms.
Measurements of $\overline{p_{\kappa}'}$ show a clear separation between the distinguishable and bosonic visible behaviors (Fig.~\ref{fig:4}a).
We expect that the main source of distinguishability in our experiment is due to thermal motional excitations normal to the lattice~(see Methods and SM Sec.~\ref{sup:budget}). This motional DOF is well-approximated by a harmonic oscillator with motional quantum numbers $n_{\rm ax}$, where atoms that possess different values of $n_{\rm ax}$ are distinguishable.
Our measurements of $\overline{p_{\kappa}'}$ are consistent with a thermal occupation of $\langle n_{\rm ax} \rangle = 0$, corresponding to the fully indistinguishable case, and inconsistent with significantly higher thermal occupation ($\langle n_{\rm ax} \rangle \gtrsim 0.167$).

Due to parity projection, $\overline{p_{\kappa}'}$ is closely related to the probability an atom is detected after the quantum walk dynamics, and is akin to standard measurements performed in the photonics community using click detectors~\cite{zhong_quantum_2020,deng_gaussian_2023}.
It is therefore important to calibrate any errors that lead to a modification of the observed atom number, including single-particle loss and certain kinds of detection errors, and include them in our simulations (see Methods and SM Sec.~\ref{sup:genbunchsim}).
This is in contrast to the measurements in Fig.~\ref{fig:2} and the Extended Data, which are insensitive to such errors due to postselection.
Despite being sensitive to calibration errors, the agreement between our measurements of $\overline{p_{\kappa}'}$ and low temperature simulations suggest that the indistinguishability measured via few particle calibrations is not noticeably degraded when scaling our experiments up to more particles.
This motivates experiments with large ensembles of atoms whose behavior we are unable to simulate, which we now discuss.


The largest input patterns we study in this work contain 180 atoms (Fig.~\ref{fig:4}bc).
Although these patterns can be prepared with no defects, in subsequent measurements we no longer enforce perfect rearrangement to avoid incurring significant overhead in the number of required experimental trials.
However, because we image the atoms after rearrangement and before their evolution, we can identify the locations of any defects.
This results in a version of scattershot boson sampling~\cite{lund_boson_2014}, but with much less variation in input states than is typical~\cite{bentivegna_experimental_2015}.
Based on the few-particle characterizations of atom indistinguishability performed across relevant regions in the lattice~(see SM Sec.~\ref{sup:calibration}), we expect the on-demand success rate for preparing a single ground-state atom, evolving it with no loss, and detecting its position in an arbitrary site to be $\sim92\;\%$.
The dominant source of deviation from perfect boson sampling is from $5.0(2)\;\%$ atom loss due to imperfect cooling in the in-plane directions, with additional contributions from imaging, rearrangement, and distinguishability~(see Methods and SM Sec.~\ref{sup:budget}).
Given these calibrations we expect that, on an average run of the experiment, $\sim166$ of the 180 input sites are populated with identical bosons that evolve under approximately Haar-random unitary dynamics (see Methods) to $\sim 1015$ output sites in the lattice with no loss or detection errors.

The above observations suggest that the experiment is performing a difficult sampling task, but it is not feasible to directly verify that the collected samples are indeed drawn from the correct distribution.
In order to compare to simulations, we control the distinguishability of the atoms by introducing additional time labels.
For a distinguishability that is high enough to allow for classical simulation, we find good agreement between theory and measurements of both $\overline{p_{\kappa}'}$ (Fig.~\ref{fig:4}b), and of the distribution of atom survival probabilities (Fig.~\ref{fig:4}c).
We observe the expected qualitative behavior where reduced distinguishability leads to an increase in $\overline{p_{\kappa}'}$ and reduced atom survival.
The above measurements serve as indirect evidence that the experiment with no additional time labels is behaving in line with expectations for bosonic particle statistics.



It is instructive to compare our approach to pioneering experiments that study boson sampling using photons (Extended Data Tab.~\ref{etab:compare}).
In our experiment, we benefit from low loss (see SM Sec.~\ref{sup:calibration}) that is not strongly dependent on evolution time (equivalently, the depth of the applied linear optical network)~\cite{garcia-patron_simulating_2019}, high state preparation and detection fidelity, and many lattice sites (equivalently, many output modes).
However, like in previous large-scale demonstrations of boson sampling~\cite{zhong_quantum_2020, madsen_quantum_2022, deng_gaussian_2023}, we currently can only apply a restricted family of unitaries.
These unitaries possess additional structure that could, in principle, be taken advantage of in efficient classical simulations.
Haar-averaging of the unitary would remove the possibility of such simulations, and provide access to additional tests of boson sampling that rely on random matrix theory~\cite{Boixo2018}.
Based on previous demonstrations, universal control over the single particle unitary~\cite{reck_experimental_1994} can in principle be implemented using the same optical tweezers we use for atom rearrangement without introducing additional loss~\cite{young_tweezer-programmable_2022}.
However, an important open problem is to improve the efficiency of protocols for directly calibrating the applied unitary for large systems, especially since errors in the applied unitary can have dramatic consequences on the resulting interference~(see Methods and SM Sec.~\ref{sup:error_model}).

In this work, we have demonstrated a new approach to performing large-scale boson sampling that is enabled by a unique combination of tools for the rapid assembly, evolution, and detection of individual atoms in a tunnel-coupled optical lattice, as well as techniques for benchmarking the quality of state preparation and evolution in such a system.
In the future, we expect that flexible programmability of the single-particle unitary using optical tweezers will enable stronger certifications of the high-order interference at the core of boson sampling~\cite{tichy_stringent_2014, dittel_totally_2018}, as well as studies of dynamical phase transitions in sample complexity~\cite{deshpande_dynamical_2018}.
More broadly, these tools could be combined with other types of atoms, and with the controllable interactions that are readily available in atomic systems, to rapidly assemble and study interacting Hubbard models~\cite{bloch_many-body_2008, yan_two-dimensional_2022-1} for simulations of condensed matter physics~\cite{lee_doping_2006}, to perform tests of complexity in the presence of interactions~\cite{maskara_complexity_2022, zheng_efficiently_2022}, and to realize new approaches to computing~\cite{childs_universal_2013}, including in hardware-efficient fermionic architectures~\cite{gonzalez-cuadra_fermionic_2023-1}.


\droptocpage

\bibliography{references,shawn_refs}

\section{Methods}




\subsection{State preparation}
\label{m:statePrep}

State preparation in our experiment involves atom rearrangement and high-fidelity optical cooling.
The rearrangement must balance several conflicting requirements associated with working in a tightly-spaced optical lattice that is compatible with strong tunnel coupling.
In particular, our tweezers have a beam waist of 480(20)~nm in comparison to the lattice spacing of $\sim 813/\sqrt{2}$~nm.
This means that tweezers directed at a specific lattice site still have substantial overlap with adjacent sites, and so atoms trapped in those adjacent sites can experience higher loss rates, especially when the tweezers are moving.
Additionally, overlapping tweezers will beat against each other due to our use of crossed acousto-optic deflectors to project the tweezer array~\cite{norcia_microscopic_2018}.
Tweezers on adjacent lattice sites are separated in frequency by 1.95~MHz, far higher than the relevant trap frequencies of 50-200~kHz, and so 1D tweezer arrays with equal spacing to the lattice don't cause significant heating. However, because the lattice is slightly rectangular, placing a tweezer on every lattice site in 2D can result in lower beat frequencies that cause significant heating.
Our algorithm for atom rearrangement seeks to balance these concerns against the desire to minimize the total distance travelled by the atoms and the total duration of the rearrangement procedure.

The basic premise is to perform most operations in parallel with 1D arrays of tweezers~\cite{ebadi_quantum_2021,tian_parallel_2023}, and to perform as many operations as possible on a sublattice that does not lead to undesirable beat frequencies between overlapping tweezers. We choose to work with either nearest- or next-nearest-neighbor spacing along one axis, and third-nearest-neighbor spacing along the other axis.
Atoms are stochastically loaded into tweezers on this sublattice with $50-75\;\%$ filling (depending on whether or not enhanced loading is used~\cite{jenkins_ytterbium_2022-1}) and implanted into the lattice~\cite{young_tweezer-programmable_2022}. Rearrangement proceeds in four stages (Fig.~\ref{efig:1}a):

\begin{enumerate}
	\item \textbf{Pre-sorting:} If a column is loaded with more atoms than are required in the target pattern, any excess atoms are pushed to the neighboring column in a single step using a 1D tweezer array. Similarly, if the column has too few atoms the missing atoms are pulled from the adjacent column. In the rare case that this step fails, the experiment is terminated and we simply load a new ensemble of atoms.
	\item \textbf{1D rearrangement:} Each column is then rearranged in 1D, with excess atoms pushed to the edges of the array.
	\item \textbf{Filtering:} A subset of atoms are transferred back into a 2D tweezer array that only addresses the correct sublattice, and the lattice potential is extinguished. Laser light that addresses the $\rm{{^1S_0} \leftrightarrow {^3P_1}}$ transition is applied to resonantly blow away the excess atoms while optically cooling the tweezer-trapped atoms. The remaining atoms are subsequently transferred back into the lattice.
	\item \textbf{Compression:} Optionally, the columns of rearranged atoms can be translated closer together one column at a time. 
\end{enumerate}

The number of steps in the resulting algorithm scales as $O(\sqrt{n})$, where $n$ is the number of atoms in the target pattern.
For target patterns with similar density to the loaded sublattice, the mean distance travelled by each atom is $O(1)$, and the runtime of the algorithm scales like $O(\sqrt{n})$.
For dense target patterns the worst-case scaling of the mean distance travelled is $O(\sqrt{n})$ leading to a runtime of $O(n)$. 

Throughout rearrangement the lattices are held at a constant depth of $U_{\rm ax}/\hbar = 2\pi\times 2.4$~MHz for the axial lattice, and $U_{\rm 2D}/\hbar = 2\pi\times 1.7$~MHz for the 2D lattice.
Resolved sideband cooling is continuously applied to the lattice-trapped atoms.
To move a given set of atoms, tweezers are ramped on and off over $\rm 60~\mu s$ to a depth of $U_{\rm tw}/\hbar \simeq 2\pi\times 30$~MHz and moved with a constant speed of $\rm 26~\mu m/ms$. Rearrangement occurs under magic conditions in the lattice for the $\rm{{^1S_0} \leftrightarrow {^3P_1}}$ transition~\cite{young_tweezer-programmable_2022}, and so the tweezer-trapped atoms are shifted out of resonance from the cooling light while they are being translated. Because the appropriate moves in our algorithm can be computed efficiently, and due to the architecture of our tweezer control system~\cite{young_half-minute-scale_2020}, analyzing the atom images and programming the control system takes only $\lesssim5$~ms for target patterns containing up to 270 atoms. However, there is currently a technical delay of 110~ms between taking the image of the stochastically loaded atoms and initiating rearrangement due to the time it takes to extract the image data from the camera. This could be addressed in the future by performing image processing on the same field-programmable gate array (FPGA) hardware used for atom rearrangement~\cite{wang_accelerating_2023}.

For the 180 atom patterns in this work, rearrangement is performed in $\sim30$~ms. The per-atom success rate for this rearrangement can be as high as $99.5\;\%$, and is primarily limited by imaging fidelity and loss.
However, due to drifts on the experiment, the typical per-atom success rate is $98\;\%$ for the experiments appearing in this work.

Once the success of this rearrangement is verified via a second image, the atoms are optically cooled to near their 3D motional ground state. This cooling is composed of 120 pulses which alternate between cooling pulses with a duration of $\rm 200~\mu s$ on two nearly-orthogonal ``radial'' axes in the plane of the 2D lattice separated by $\rm 400~\mu s$ cooling pulses on the ``axial'' out of plane axis. Each cooling pulse is separated by a delay of $\rm 200~\mu s$ leading to an overall cooling sequence that is 60~ms in duration. This sequence is optimized for high-fidelity cooling of the axial direction at the cost of slightly worse radial cooling. We choose to make this trade-off because we are able to postselect for perfect cooling in the radial directions. Specifically, for our experimental parameters, atoms that occupy higher bands in the 2D lattice are lost during tunneling~\cite{young_tweezer-programmable_2022}.
Based on a master equation calculation (Fig.~\ref{efig:1}b), we expect this cooling to result in a 3D motional ground state occupation of $96.97\;\%$, which is consistent with the expectation based on an approximate analytical calculation~\cite{eschner_laser_2003}. The expected axial motional ground state occupation of $99.58\;\%$ leads to an expected indistinguishability of $99.27\;\%$, in agreement with our experimental estimate of $99.5^{+0.5}_{-1.6}\;\%$. The expected combined ground state occupation of $97.37\;\%$ in the radial directions leads to $2.63\;\%$ loss, which explains a significant fraction of our average observed loss of $5.0(2)\;\%$.
The remaining loss is likely due to imperfect adiabaticity of ramps in the lattice depth, as well as imperfect overlap between the Wannier functions of the atoms in the lattice before and after the lattice is quenched to the conditions used for tunneling.
Note that because both cooling and imaging involve the spontaneous emission of photons from atoms occupying different lattice sites, our state preparation is expected to be perfectly dephasing. This rules out error models for boson sampling like the mean field sampler that rely on coherences between different input modes~\cite{tichy_stringent_2014}.

\subsection{Image analysis and feedback}
\label{m:image}

Similar to previous works, our image analysis involves multiplying raw images by a discrete set of masks corresponding to the known locations of sites in the lattice, and thresholding the results to identify the presence or absence of an atom in a given site~\cite{young_tweezer-programmable_2022}.
However, to improve the fidelity of this procedure, we optimized the applied masks by training a single-layer neural network on simulated data corresponding to $30\;\%$ random filling of the lattice.
The resulting masks are similar to the ones we previously employed based on the measured point spread function of our imaging system, but with a slight negative bias on adjacent lattice sites that reduces errors due to leakage light from one lattice site to another.
Although a deep neural network can result in better performance~\cite{impertro_unsupervised_2023}, these performance gains are marginal in our setup due to the already high imaging resolution in comparison to the lattice spacing. This being the case, we opt to use a single-layer neural network to ease the characterization of errors and image fidelity.

The resulting analysis yields a combined imaging loss and infidelity of $p_{10} = 0.002(1)$ for a calibration data set, however, $p_{10}$ can fluctuate day to day by $\sim0.002$.
$p_{10}$ can be interpreted as the probability of a false negative, namely that a site which contains an atom is incorrectly identified as being empty. 
The probability of a false positive (where an empty site is incorrectly identified as containing an atom) is much lower, with a value of $p_{01}\sim10^{-5}$.
For experiments with very low density (for example when one atom occupies an analysis region containing $\sim1015$ sites) the effect of false positives can become comparable to or larger than the effect of false negatives when trying to correctly identify the presence and position of the atom after the quantum walk dynamics.

All imaging errors are quadratically suppressed when postselecting for perfect rearrangement and no lost or extra observed atoms, and so measurements involving such postselection are negligibly impacted by imaging errors.
For experiments in which we do not postselect on the number of atoms remaining after the quantum walk dynamics (namely the experiments appearing in Fig.~\ref{fig:4}) imaging errors can play a role.
Specifically, for the atom and mode numbers relevant to Fig.~\ref{fig:4}, the combined effects of false positives and negatives lead to overall error rates of $\sim1\;\%$ in our ability to correctly identify the presence and location of a given atom.
In experiments without postselection on the final atom number, we calibrate the effect of such imaging errors and include them in our simulations (see SM Sec.~\ref{sup:genbunchsim}).

One important consideration in our experiment is the possibility that the position of the lattice drifts relative to the tweezer array, and to the masks used in image analysis, which can occur on a timescale of $\sim30$ minutes.
To correct this, we use the images taken before and after rearrangement (but not after the atoms are allowed to propagate through the lattice) to identify the positions of the lattice-trapped atoms relative to the imaging system, and thus also relative to the optical tweezers.
This information can be used to correct any drifts by adjusting the tweezer and mask positions on subsequent runs of the experiment.
Because the repetition rate of the experiment is $\sim1$~Hz, these corrections can be made much faster than drifts can occur.
The tweezer positions can drift relative to the imaging system on a timescale of several hours, but this is readily corrected by taking images of atoms trapped in the tweezers~\cite{norcia_microscopic_2018} and adjusting their positions to match the lattice.

For data sets in which we simulate $n$ partially or fully distinguishable particles by combining multiple runs with fewer than $n$ particles, we compensate for parity projection during imaging by summing the resulting processed images and taking the result $\mathrm{mod}(2)$.

\subsection{Computing atom indistinguishability}
\label{m:computing_particle_indistinguishability}
The goal of our Hong-Ou-Mandel (HOM) experiments is to measure the indistinguishability of a pair of atoms.
To do so, we measure the probability that two nominally indistinguishable particles arrive in disjoint subsets of sites, and compare this to the corresponding probability for distinguishable particles.
We now show that the ratio of these two quantities gives a lower bound on the indistinguishability of the atoms.

Our model is that we have linear optical evolution on atoms with an extra degree of freedom that evolves independently of the visible degree of freedom. 
Then, the probability that two partially distinguishable atoms start in the sites $k, l$ with $k \neq l$ and end in sites $i, j$ is
\begin{align}
  p^{\text{partial}}(ij|kl) &= |U_{i,k}|^2|U_{j, l}|^2+|U_{i,l}|^2|U_{j,
  k}|^2\nonumber\\ 
  &+ 2\mathcal{J} \operatorname{Re}\left( U_{i, k}U_{j, l}U_{j, k}^*U_{i, l}^* \right) \\
  \text{if $i \neq j$}\nonumber\\
  p^{\text{partial}}(ii|kl) &= (1+\mathcal{J})|U_{i,k}|^2|U_{i,l}|^2
  \label{eq:partial}
  \\
  \text{otherwise}\nonumber
\end{align}
where the indistinguishability $\mathcal{J}$ is the Hilbert-Schmidt inner product of the two single-particle density matrices on the hidden degrees of freedom, and $U$ is the single particle unitary.
Correspondingly, distinguishable atoms have the distribution
\begin{align}
  p^{\text{dist}}(ij|kl) &= |U_{i,k}|^2|U_{j, l}|^2+|U_{i,l}|^2|U_{j,
  k}|^2  \qquad
  \text{if $i \neq j$}\\
  p^{\text{dist}}(ii|kl) &= |U_{i,k}|^2|U_{i,l}|^2\qquad \text{otherwise}
  \label{eq:dist}
\end{align}
In the experiment we expect that the probability of coincidence of atoms on any two particular sites is small, and therefore estimation of such a probability is difficult.
Thus, it is useful to be able to bundle many sites together.

So, let $S_1$ and $S_2$ be disjoint sets of sites.
The total probability that we prepare partially distinguishable atoms, and see one atom each in $S_1, S_2$ is
\begin{align}
  p^{\text{partial}}_{S_1, S_2} = \sum_{\substack{i \in S_1\\j\in S_2}}p^{\text{partial}}(ij|kl)
  \label{eq:totalcoincidencepartial}
\end{align}
and the corresponding probability for distinguishable atoms is
\begin{align}
 p^{\text{dist}}_{S_1, S_2}&=  \sum_{\substack{i \in S_1\\j\in S_2}}p^{\text{dist}}(ij|kl)
  \label{eq:totalcoincidencedist}
\end{align}
so their ratio is
\begin{align}
  Q^{\text{HOM}}_{S_1, S_2} \coloneqq 1+ \mathcal{J} \underbrace{\frac{2\sum_{\substack{i \in S_1\\j\in S_2}}\Re( U_{i, k}U_{j, l}U_{j, k}^*U_{i, l}^* )}{ \sum_{\substack{i \in S_1\\j\in S_2}}\left(|U_{i,k}|^2|U_{j, l}|^2+|U_{i,l}|^2|U_{j, k}|^2\right)}}_{\coloneqq -\tau(S_1, S_2)}
  \label{eq:subsethom}
\end{align}
Note that the ratio $-\tau(S_1, S_2) \ge -1$.
We thus have the inequality
\begin{align}
  Q^{\text{HOM}}_{S_1, S_2} \ge 1- \mathcal{J}   
  \label{eq:subsethomineq}
\end{align}
When equality holds so $\tau(S_1, S_2) = 1$, we say that $S_1$ and $S_2$ satisfy the balanced condition, in analogy with the case of a balanced beamsplitter.

\subsubsection{HOM with loss and parity projection}
In our experiment, we are interested in the case that $S_1 = C_k$ and $S_2 = C_l$, where $C_k$ is the column of lattice sites that includes site $k$, and similarly for $C_l$.
It remains to construct an estimator of $Q^{\text{HOM}}_{C_k, C_l}$.
We will construct it from separate estimates of $p^{\text{dist}}_{C_k, C_l}$ and $p^{\text{partial}}_{C_k, C_l}$.

To estimate $p^{\text{dist}}_{C_k, C_l}$ (called $p^D_{11}$ in the main text), we can directly use the single particle data.
Denoting the event that the particle is not lost as $\neg \lambda$, we can write the coincidence probability for distinguishable particles as
\begin{align}
  p^{\text{dist}}_{C_k, C_l} &= p(C_k | k, \neg \lambda)p(C_l | l, \neg \lambda) + p(C_k | l, \neg \lambda)p(C_l | k, \neg \lambda)
  \label{eq:distcoinc}
\end{align}
Since the right hand side is a multilinear polynomial of single-particle probabilities, the plug-in estimator is unbiased.

Now we construct an estimator of $p^{\text{partial}}_{C_k, C_l}$ (called $p^B_{11}$ in the main text).
The probability that the prepared nominally indistinguishable particles end up on the same columns as the ones they start on is
\begin{align}
  p^{\text{obs}}_{C_k, C_l} = p(\nu_2 | kl)p^{\text{partial}}_{C_k, C_l}
  \label{eq:pb11}
\end{align}
where the prepared sites are named $k$ and $l$, and the event $\nu_2$ is that neither of the two particles is lost due to a single particle loss event.
Due to the effect of parity projection, we cannot measure $p(\nu_2|kl)$ without further assumptions.
In particular, we assume that the loss acts independently and identically on each site.
Then defining $\beta$ to be the event that we observe one particle in the output, the probability of $\beta$ occurring is
\begin{align}
  p(\beta|kl) = 2(1-p_\lambda)p_\lambda
  \label{eq:iidonesurvive}
\end{align}
where $p_\lambda$ is the probability that a single particle is lost due to a single particle loss event.
So we can solve for the loss probability,
\begin{align}
  p_\lambda = \frac{1 - \sqrt{1- 2 p(\beta|kl)}}{2}
  \label{eq:quadsol}
\end{align}
and compute the probability that neither particle is lost due to a single-particle loss event:
\begin{align}
  p(\nu_2|kl) &= (1-p_\lambda)^2 
  \label{eq:neitherlost}
\end{align}
Then we can construct a plug-in estimate of $Q_{C_k, C_l}^{\text{HOM}}$ from those of $p^{\text{partial}}_{C_k, C_l}$ and $p^{\text{dist}}_{C_k, C_l}$, and apply the delta method~\cite{Shao2003} to obtain a first-order unbiased estimate of $Q_{C_k, C_l}^{\text{HOM}}$.
To construct a confidence interval, we construct 1000 bootstrap estimates of $Q_{C_k, C_l}^{\text{HOM}}$ and apply the bias-corrected percentile method~\cite{Efron1994}.

From a calibration of the unitary (see SM Sec.~\ref{sup:calibration}), we can calculate $\tau(C_k, C_l)$. 
Then we can extract the indistinguishability from
\begin{align}
  \mathcal{J} = \frac{1}{\tau(C_k, C_l)}\left(1-Q_{C_k, C_l}^{\text{HOM}}\right)
  \label{eq:jestimate}
\end{align}
We expect that our uncertainty in our calibration in $\tau$ is negligible compared to the statistical fluctuation in our estimate of $Q_{C_k, C_l}^{\text{HOM}}$ (see SM Sect~\ref{sup:error_model}), so we ignored this effect when calculating a confidence interval for $\mathcal{J}$.
The confidence interval was constructed through the bias-corrected percentile method, using 1000 bootstraps.
When constructing our point estimate and our confidence interval, the estimate may go above 1 since $\tau \le 1$.
To account for this, after constructing the bootstrap confidence interval, we clip the upper end to be equal to 1 if it is larger than 1.
Similarly, we apply the same procedure to the point estimate.

\subsection{Binning}

There is flexibility in how to partition between visible and hidden DOFs (Fig.~\ref{fig:1}b), which we take advantage of in tests of indistinguishability involving up to 8 atoms.
In an ideal square lattice, $H$ takes the form $H_{x}\otimes \mathbb{1}_y + \mathbb{1}_x \otimes H_{y}$ where $x$ and $y$ denote the spatial coordinates of the lattice sites, and in this case one can consider one of the spatial coordinates as hidden.
For example, if we ignore the $y$ coordinate of the atoms the resulting visible behavior is that of a 1D multiparticle quantum walk along $x$ (Fig.~\ref{fig:1}cd).
We refer to this as ``binning,'' since this is accomplished by summing the observed numbers of atoms along the $y$ axis in the final image.
Binning is convenient because it allows us to operate in a regime where it is rare that there is more than one atom on a given lattice site, minimizing the effect of parity projection, and providing effectively number-resolved measurements of the visible sites.
Note that the above form of $H$ is weakly violated in our lattice due to higher order tunneling terms (e.g. diagonal tunneling) and small non-factorizable variations in the nearest-neighbor tunneling terms.
These violations are accounted for in our simulations, and do not significantly affect our results.

\subsection{Clouding}

The clouding probability is defined as the probability that all atoms end up on the same half of the array for a 1D quantum walk~\cite{carolan_experimental_2014} (Fig.~\ref{fig:2}bde).
Similar to bunching, we normalize the clouding probability of distinguishable particles by the probability of full survival for bosonic particles to account for the effect of parity projection (see SM Sec.~\ref{sup:distinguishableconfint_bunchcloud}).
Note that, unlike full bunching, the enhancement of clouding for identical bosons in comparison to distinguishable particles is strongly dependent on both the system evolution, and the specific input state.
For example, preparing atoms at next-nearest-neighbor spacing in the lattice can make the difference in clouding for bosonic and distinguishable atoms almost zero, as confirmed by measurements of two and three atoms (Fig.~\ref{fig:2}ad, Extended Data Fig.~\ref{efig:3}).

\subsection{Generalized bunching}

The generalized bunching probability $p_\kappa$ that all $n$ atoms appear in an arbitrary subset $\kappa$ of sites can serve as a useful quantity for benchmarking the performance of a boson sampler~\cite{shchesnovich_universality_2016, seron_efficient_2022}.
For most unitaries, including those applied in this work, numerical calculations indicate that $p_\kappa$ is maximized by bosonic particle statistics in comparison to other particle statistics.
Interestingly, this is not true in certain fine-tuned cases~\cite{seron_boson_2023, pioge_enhanced_2023}, which would be interesting to explore in future experiments with more control over the applied unitary.

The above generalization of bunching helps to unify our earlier measurements, where binning columns of sites and measuring coincidences, full bunching, or clouding simply correspond to specific choices of $\kappa$ (Fig.~\ref{fig:1}b).
For appropriate selections of the size of $\kappa$, the difference in generalized bunching probabilities of bosonic behavior in comparison to other behaviors is expected to converge in a number of measurements that is polynomial in $n$~\cite{shchesnovich_universality_2016, shchesnovich_distinguishing_2021}.
Specifically, we choose $|\kappa| = \lfloor m-m/n\rceil$ (where $\lfloor\cdot\rceil$ denotes rounding to the nearest integer), with $m = 500$ for Fig.~\ref{fig:4}a, and $m = 1015$ for Fig.~\ref{fig:4}b

We cannot directly measure $p_\kappa$, and instead measure the probability $p_{\kappa}'$ that all observed atoms on a given run of the experiment appear in the set of sites $\kappa$.
$p_{\kappa}'$ differs from $p_\kappa$ because even-numbered occupancy of a site not contained in $\kappa$ contributes to a ``successful'' event where all remaining atoms after parity projection appear within $\kappa$.
Although we do not claim that $p_{\kappa}'$ is maximized by perfectly bosonic visible behavior for any input state, $\kappa$, and $U$, we find in numerics that $p_{\kappa}'$ still serves as a useful observable: it converges in a reasonable number of measurements, and distinguishes between a family of experimentally relevant models for the hidden DOFs (discussed in the main text).
To avoid concerns of biasing in our choice of $\kappa$, we average $p_{\kappa}'$ over all choices of a given size $k \coloneqq |\kappa|$ to compute the quantity $\overline{p_{\kappa}'}$, which we refer to as the ``modified generalized bunching probability'' in the main text.
Note that the main contribution to $\overline{p_{\kappa}'}$ is bunching and the resulting loss as a result of parity projection.
As a result, we do not postselect on the number of surviving atoms after the quantum walk dynamics in our measurements of $\overline{p_{\kappa}'}$.

Although there are many selections of $\kappa$ that must be averaged over, $\overline{p_{\kappa}'}$ can be estimated efficiently from the observations via the following combinatorial argument:
Let $[m] = \left\{ 1, \ldots, m \right\}$ be shorthand for the set of output sites. Let $S \subseteq_{k} [m]$
denote a subset of sites of size $k$. 
Let $G$ be the random variable denoting the site occupation of the output, with $g$ being a single sample of that random variable.
Then, the average probability that all $n$ particles arrive in a set of size $k$ is
\begin{align}
  \overline{p_\kappa'} &= {m \choose k}^{-1} \sum_{S \subseteq_{k} [m]}\operatorname{Pr}(G \subseteq S) \\
  &=  {m \choose k}^{-1} \sum_{g}\operatorname{Pr}(g) \sum_{S \subseteq_{k} [m]} \mathbb{I}(g \subseteq S)\\
  &= {m\choose k}^{-1}\sum_{g}\operatorname{Pr}(g) {m - \#(g) \choose k - \#(g)}
  \label{eq:fullsum}
\end{align}
where $\mathbb{I}$ is the indicator function that is $1$ when its argument is true, and $0$ otherwise, and $\#(g)$ is the number of nonzero entries of $g$, and the sum over $g$ ranges over all possible mode occupations of the output.
Since the expression in Eq.~\ref{eq:fullsum} is a linear combination of probabilities of outcomes, we can weight the corresponding frequencies by the coefficients appearing in the sum to estimate the quantity $\overline{p_{\kappa}'}$.

\subsection{Quantum walk dynamics}
The single-particle Hamiltonian governing the evolution of atoms in the lattice is:
\begin{align}
  \label{eq:Ham}
  H &= -\sum_{\langle i,j \rangle} J_{ij} \left(\ketbra{i}{j} + \ketbra{j}{i}\right) - \sum_{i} V_i \ketbra{i}{i},
\end{align}
where $\ket{i}$ denotes occupation of the lattice site $i$, and $\langle i,j \rangle$ denotes all pairs of sites in the lattice.
$V_i$ denotes a position-dependent potential that captures the harmonic confinement imposed by the lattice beams and, in principle, can be adjusted using the optical tweezers~\cite{young_tweezer-programmable_2022}.
$J_{ij}$ is the strength of the tunnel coupling between sites $i$ and $j$. Nearest-neighbor tunneling dominates with an energy of $J_{ij}/\hbar \simeq J/\hbar = 2\pi\times 119$~Hz, however, small contributions from diagonal and next-nearest-neighbor tunneling are also included in our simulations (see SM Sec.~\ref{sup:calibration})~\cite{young_tweezer-programmable_2022}.
Evolution under $H$ results in a quantum walk described by an $m\times m$ single-particle unitary $U = e^{-i H t/\hbar}$, where $m$ is the number of sites in the lattice.

As mentioned in the text, some loss occurs during the evolution as a result of imperfect state preparation, but we do not observe this loss to depend on $t$ (although we do expect additional loss effects to play a role on timescales that are very long in comparison to the dynamics, see SM Sec.~\ref{sup:calibration}). 
This contrasts with previous demonstrations of boson sampling where the survival probability decays exponentially with evolution time (or equivalently the depth of the linear optical circuit), which can be exploited in classical simulation methods~\cite{garcia-patron_simulating_2019,Oszmaniec_2018,PhysRevA.104.022407,oh_classical_2023}.

For many of our measurements with 1D nearest-neighbor input patterns, we use an evolution time of $t = (n-1) t_{\rm HOM} $ such that all input sites are approximately uniformly coupled to each other after the quantum walk dynamics~\cite{muraleedharan_quantum_2019}.
At an evolution time of 6.45~ms, as applies to the data appearing in Fig.~\ref{fig:4}, $U$ does not significantly couple all 180 input sites to all $\sim1015$ output sites.
This is primarily due to the finite size of the lattice beams and resulting harmonic confinement (see Extended Data Fig.~\ref{efig:2}ab and SM Sec.~\ref{sup:calibration}).
This harmonic confinement is negligible in a $15\times15$-site region near the center of the lattice, in which the distribution of the norm-square of the elements in $U$ is well-captured by the Porter-Thomas distribution for 385 outputs, indicating behavior that shares features with a Haar-random unitary (Extended Data Fig.~\ref{efig:2}cd).
Over the full $1015$-site region considered in this work, $U$ couples each site to an average of $83$ of the $180$ input sites with an amplitude of $\ge\!10^{-3}$.
Some sites are coupled by $\ge\!10^{-3}$ to as many as $156$ input sites. 







\subsection{From single- to many-particle dynamics}

Adopting the first-quantized treatment typically used in discussions of boson sampling~\cite{aaronson_computational_2011}: consider $n$ indistinguishable, bosonic, and non-interacting atoms occupying input sites $\vec{j} = (j_1,\dots,j_n)$, with $j_1 \le \cdots \le j_n$, where $j_l \in \{1, \dots, m\}$, and $m$ is the total number of available sites in the lattice.
Each atom undergoes a quantum walk described by an $m\times m$ single-particle unitary $U = e^{-i H t/\hbar}$, where $t$ is the evolution time.
To calculate the probability $P^{\rm B}(\vec{k}| \vec{j}, U)$ of observing a specific output pattern $\vec{k} = (k_1,\dots,k_n)$ after this evolution, one must sum over all permutations of atom labels on the input going to atom labels on the output.
This treatment is equivalent to properly symmetrizing the state and evolution associated with these atoms, and yields~\cite{aaronson_computational_2011}:

\begin{equation}
  P^{\rm B}(\vec{k}| \vec{j}, U) = \frac{1}{\vec{k}!}|\mathrm{Perm}(U_{\vec{k}, \vec{j}})|^2
	\label{eq:bosonprob}
\end{equation}

\noindent where $\smash{U_{\vec{k}, \vec{j}}}$ is the $n\times n$ submatrix of $U$ that contains only the rows corresponding to sites $\vec{k}$ (including any duplicates), and columns corresponding to $\vec{j}$ (which contains no duplicates in our experiments since the lattice is initialized with at most one atom per site).
The normalization constant $\vec{k}!$ is more conveniently expressed in the site occupation basis, $\vec{k}' = (k^{\prime}_1,\dots,k^{\prime}_m)$, where $k^{\prime}_i$ counts the number of atoms occupying a given site $i \in \{1, \dots, m\}$.
Then we define $\vec{k}! = \prod_{i=1}^{m}k^{\prime}_i!$.
The equivalent calculation for distinguishable atoms yields $P^{\rm D}(\vec{k}| \vec{j}, U) = \mathrm{Perm}(|U_{\vec{k}, \vec{j}}|^2)/\vec{k}!$.
Notice that since the single-particle distributions are given by $|U_{\vec{k}, \vec{j}}|^2$, we can simulate the distinguishable situation by combining separate single particle measurements (as applies in the case of time-labeled distinguishable data throughout this work).

\subsection{Simulations}
For up to 3 particles, the simulations in this work involve exactly solving for the full output distribution by evaluating the permanent in Eq.~\ref{eq:bosonprob} (or the corresponding expression for distinguishable particles) for all possible outputs via Glynn's formula.
For larger particle numbers, we follow the approach of Clifford and Clifford~\cite{clifford_classical_2018} to sparsely sample from the full output distribution.

The simulations involving thermal occupation of the motional degree of freedom normal to the lattice in Fig.~\ref{fig:4}a assume that the evolution of this hidden degree of freedom is independent of the visible evolution of the atoms, and dephased by our state preparation.
As a result, the evolution of atoms with a mixed motional state can be simulated by assigning a specific motional state drawn from the appropriate thermal distribution to each atom in a given simulated sample.
For each subset of atoms that share a motional state we draw a sample using the approach of Clifford and Clifford~\cite{clifford_classical_2018}, and combine these samples into a single sample of partially distinguishable atoms.
The effect of loss and detection errors are simulated incoherently, and applied after these samples are generated, as is parity projection (see SM.~\ref{sup:genbunchsim}).
The above simulations indicate that the thermal model for distinguishability leads to expected measurements of modified generalized bunching that monotonically interpolate from the distinguishable to the indistinguishable bosonic case as the temperature is reduced.

\subsection{Characterizing the single particle unitary}
As discussed in the main text, we characterize the dynamics of the atoms in two ways. 
The first is the spectroscopic characterization (see SM Sec.~\ref{sup:calibration}), and the second is a maximum likelihood (ML) procedure to fit the single-particle unitary that determine the dynamics, from one- and two-particle data.
Here we describe the ML procedure and our method of determining its performance.

First we describe the model used to compute the likelihoods.
Our model describes atoms that are subjected to single particle loss, undergo tunneling dynamics, then are measured via parity projection.
Since we are only inferring some of the parameters of the unitary, it suffices to use a restricted model that only uses the entries of the single particle unitary $U$ that describe scattering from the input sites $I$ (highlighted in red in Fig.~\ref{fig:3}) to the output sites $S$ (green crosses in Fig.~\ref{fig:3}).
The model for the single particle distribution is
\begin{align}
  p_{U, p_\lambda}(s|i) &= (1-p_\lambda) |U_{si}|^2 \qquad \text{if $s \in S$}\\
  p_{U, p_\lambda}(\tau|i) &= (1-p_\lambda)\left(1-\sum_{s \in S}|U_{si}|^2\right) \\
  p_{U, p_\lambda}(\emptyset|i) &= p_\lambda.
  \label{eq:singledistn}
\end{align}
Here, $\tau$ is the event that the atom arrived outside of $S$, $\emptyset$ is the event that the particle was lost, $i \in I$ is the initial site, $p_\lambda$ is a parameter describing the single particle loss, and $U_{si}$ is the parameter describing the amplitude for one particle to start in $i$ and end in $s$.
The model for the two particle distribution is
\begin{align}
  p_{U, p_\lambda, \mathcal{\mathcal{J}}}(s, s'|i, j) &= (1-p_\lambda)^2 p_{U, \mathcal{J}}^\text{partial}(s, s'|i, j) \nonumber\\ &\text{for $\{s, s'\} \in P_2(S)$}\\
  p_{U, p_\lambda, \mathcal{J}}(s|i, j) &= p_\lambda (1-p_\lambda)(|U_{si}|^2 + |U_{sj}|^2) \nonumber\\ &\text{for $s\in S$}\\
  p_{U, p_\lambda, \mathcal{J}}(\zeta|i, j) &= 1-\sum_{\{s, s'\} \in P_2(S)}p(s, s'|i, j) - \sum_{s \in S}p(s|i, j)
  \label{eq:twodistn}
\end{align}
Here, the set $P_2(S)$ is of sets of pairs of elements of $S$, the event $\zeta$ is the event that it was not the case that all surviving particles arrived in $S$, and $i, j \in I$ are the initial sites.
Finally, the probability that lossless, partially distinguishable atoms start in sites $i, j$ and arrive at distinct sites $s, s' \in S$ is
\begin{align}
  p^{\text{partial}}_{U, \mathcal{J}}(s, s'|i, j) &= |U_{s,i}|^2|U_{s', j}|^2+|U_{s,j}|^2|U_{s', i}|^2\nonumber\\ 
  &+ 2\mathcal{J} \operatorname{Re}\left( U_{s, i}U_{s', j}U_{s, j}^*U_{s', i}^* \right) \\
  \label{eq:losslesspartial}
\end{align}
where $\mathcal{J}$ is the indistinguishability of the two atoms.

The parameters $p_\lambda$ and $U$ are the parameters that we wish to infer, while the parameter $\mathcal{J}$ is obtained from separate calibration data, as discussed in the previous section.
To simplify the inference procedure, we first infer $p_\lambda$ using only the single particle data, then with $p_\lambda$ fixed, we run the quasi-Newton L-BFGS optimizer as implemented in \textsc{pytorch} to maximize the log-likelihood of the data with respect to $U$.
It remains to specify a parameterization of $U$.

Let $M$ be the submatrix of $U$ that is the intersection of the columns specified by $I$ and the rows specified by $S$.
Then since $M$ is only $|S| \times |I|$, we wish to find a parameterization of it that does not require too many more parameters, while at the same time respecting the constraint that it is a submatrix of a unitary.
To accomplish this, we specify $M$ as the $|S| \times |I|$ top-left submatrix of a $(|S|+|I|) \times (|S|+|I|)$ unitary matrix $V$.
We parameterize the matrix $V$ by specifying its anti-Hermitian matrix logarithm $iH$, so $V = e^{i H}$.
We parameterize $H$ by writing it as a (real) linear combination of the generalized Gell-Mann matrices.
To define the $d \times d$ Gell-Mann matrices, first let $k, l, q, j \in \left\{ 1, \ldots, d \right\}$.
Then the $k, l$th Gell-Mann matrix $B_{kl}$ is defined by
\begin{align}
  (B_{kl})_{qj} &= \frac{1}{\sqrt{2}}(\delta_{qk}\delta_{jl} + \delta_{ql}\delta_{jk})\qquad\text{if $k < l$}\\
  (B_{kl})_{qj} &= \frac{1}{\sqrt{2}}(i\delta_{qk}\delta_{jl}  -i\delta_{ql}\delta_{jk})\qquad\text{if $k > l$}
\end{align}
and for $k = l < d$, we have
\begin{align}
  (B_{kk})_{qj} &= \frac{1}{\sqrt{k(k+1)}}\delta_{qj} \qquad\text{for $q, j \le k$}\\
  (B_{kk})_{(k+1),(k+1)} &= \frac{-k}{\sqrt{k(k+1)}}\\
 (B_{kk})_{qj}&= 0 \qquad \text{else}
  \label{eq:gellmann}
\end{align}
and finally, we have
\begin{align}
  B_{d, d} &= \frac{1}{\sqrt{d}}\mathbb{1}_{d}
  \label{eq:lastgellmann}
\end{align}
The Gell-Mann matrices $\{B_{ij}\}$ are a basis of Hermitian matrices that are orthonormal with respect to the Hilbert-Schmidt inner product.
Thus, we can specify $H$ in terms of its coefficients $c_{ij}$ in the basis of $(|S|+|I|) \times (|S|+|I|)$ Gell-Mann matrices, so $H = \sum_{ij}c_{ij}B_{ij}$.
This gives us a $(|S|+|I|)^2$ dimensional parameter space without a boundary.

To specify the initial point of the algorithm, we start with a model $M_0$ of $M$ as computed from the spectroscopic calibration (see SM Sec.~\ref{sup:calibration}), compute an isometric completion $W_0$ of it, expressed in block form by $W_0^\dagger = \left(M_0, \sqrt{\mathbb{1} - M_0M_0^\dagger}\right)$, then compute a unitary completion $V_0^\dagger$ of $W_0$ by appending an orthonormal basis of the nullspace of $W_0W_0^\dagger$ as columns.
Then, if the eigenvalues of $V_0$ are $e^{i\phi_k}$ where each $\phi_k \in (-\pi, \pi]$ for $k \in \left\{ 1, \ldots, |I|+|S| \right\}$, and $Q$ diagonalizes $V_0$, we construct an initial Hermitian matrix $H_0$ from $H_0 = Q\; \mathrm{diag}(\phi_1, \ldots, \phi_{n+m}) \;Q^\dagger$.
Since the Gell-Mann matrices are orthonormal, we can then extract the initial coefficients $(c_0)_{ij}$ from $(c_0)_{ij} = \Tr(H_0 B_{ij})$.

Having specified the initial point of the L-BFGS optimizer, we can run it to maximize the log-likelihood of the data.
The data has seven measurement settings, consisting of four single-particle settings, and three two-particle settings, as shown in Fig.~\ref{fig:3}, and the numbers of experiments performed in each setting are given in SM Sec.~\ref{sup:stats}.
Unfortunately during the optimization, sometimes the parameters run off to very large values, leading to numerical instability.
In these cases, we simply restart the algorithm with slightly adjusted parameters, by adding shifts $s_{ij}$ to the initial parameters $(c_0)_{ij}$ drawn from independent Gaussians of mean zero and standard deviation $0.1$.

We would like to get a sense of the performance of this inference procedure, and in particular whether our calibrated model deviated from the ML estimate more than would be expected from statistical fluctuation.
The calibrated model $p_{M_0, (p_\lambda)_0, \mathcal{J}_0}$ is specified by the evolution parameters $M_0$, the loss parameter $(p_\lambda)_0$ and the indistinguishability parameter $\mathcal{J}_0$.
The evolution parameters $M_0$ are computed from the spectroscopic characterization.
The loss parameter $(p_\lambda)_0$ and indistinguishability parameter $\mathcal{J}_0$ are computed from the HOM data that is used in the main text.
The loss $(p_\lambda)_0$ is the frequency that no particles survived in the one particle preparations of the HOM data,
the indistinguishability $\mathcal{J}_0$ was computed from the method described in the previous section.

To quantify the deviation of the ML fit $(M^*, (p_\lambda)^*)$ to the calibrated model, we use the total variation distance of the implied distributions.
Specifically, we compute the total variation distances between $p_{M_0, (p_\lambda)_0, \mathcal{J}_0}(\cdot|a)$ and $p_{M^*, (p_\lambda)^*, \mathcal{J}_0}(\cdot|a)$ for each one- and two- particle input $a$, and take the maximum of the results.
We call this the maximum total variation distance (max TVD) $d(p_{M_0, (p_\lambda)_0, \mathcal{J}_0}, p_{M^*, (p_\lambda)^*, \mathcal{J}_0})$.
The max TVD has the following operational interpretation:
suppose that we are allowed to choose among the 7 measurement settings to perform a single experiment, and our task is to decide whether the parameters that describe the system are $(M_0, (p_\lambda)_0, \mathcal{J}_0)$ or $(M^*, (p_\lambda)^*, \mathcal{J}_0)$.
Then $\frac{1}{2} + \frac{1}{2} d(p_{M_0, (p_\lambda)_0, \mathcal{J}_0}, p_{M^*, (p_\lambda)^*, \mathcal{J}_0})$ is the optimal probability with which we could guess correctly.

We would like to capture the statistical variation in the max TVD.
To do so, we perform bootstrap resamples of the HOM data to obtain a bootstrap estimate $\mathcal{J}^b_i$ of the indistinguishability, which is then used to perform ML on bootstrap resampled data, to obtain the bootstrap fit parameters $(M^b_i, (p_\lambda)^b_i)$.
We then compute the histogram of values $d(p_{M_i^b, (p_\lambda)_i^b, \mathcal{J}^b_i}, p_{M^*, (p_\lambda)^*, \mathcal{J}_0})$.

We would like to compare the value of $d_{0*} \coloneqq d(p_{M_0, (p_\lambda)_0, \mathcal{J}_0}, p_{M^*, (p_\lambda)^*, \mathcal{J}_0})$ to the resulting histogram.
In Fig.~\ref{efig:2}e we show a bootstrap histogram that shows the max TVD from the point estimate to the bootstrap ML estimates.
We can see that $d_{0*}$ is slightly larger than the mean of the bootstrap distribution.
This is the expected behavior because statistical fluctuations in the calibrated model also contribute to $d_{0*}$, but a more thorough characterization of the statistical fluctuations in the calibrated model would be required to confirm this.
Also shown in Fig.~\ref{efig:2}e are the max TVDs from $p_{M_0, (p_\lambda)_0, \mathcal{J}_0}$ and $p_{M^*, (p_\lambda)^*, \mathcal{J}_0}$ to the frequencies of the data.

\subsection{Validating few-particle quantum walks}

To validate the quality of the quantum walks performed in this work, we determine
the total variation distances (TVDs) between the experimental
probability distributions and those expected from the calibrated
model. In the limit of many samples, one can estimate this by
determining the empirical TVD between the experimentally observed
frequencies and those expected according to the model.  When, as in our experiment, the
number of trials is relatively small, the
empirical TVD is expected to be biased high.  We therefore compare
the empirical TVD to that expected if one were to sample from an experimental probability
distribution that is equal to the model. The latter distribution is estimated by Monte Carlo sampling from the model distribution.

In Fig.~\ref{efig:4} we show the binned experimental frequencies and
the probability distributions of the model for \(2\) and \(3\) atom
quantum walks at evolution times of $t_{\mathrm{HOM}}$ and
$2t_{\mathrm{HOM}}$ respectively. The corresponding empirical TVDs
are \(0.046\) and \(0.109\). The expected empirical TVDs if
the experimental distributions are the same as the
models' are $0.047$ and $0.099$. The
standard deviations of the distributions of these empirical TVDs
observed by Monte Carlo sampling are \(0.009\) and \(0.006\)
respectively. This comparison shows that although the empirical TVDs
are conservatively estimated upper bounds on the true TVDs, they are
consistent with having TVDs of zero. Useful estimates require
many more experimental trials, which is challenging with our experimental cycle time of
about \(1\;\text{s}\).

Instead of estimating the TVD directly, we can estimate the
contributions of known discrepancies between the model and the
experimental implementation.  In addition to the distinguishability
and loss errors discussed in the main text, there are two kinds of
errors arising from evolution under different unitaries than
that of the calibrated model: calibration error, which leads to
a systematic discrepancy between the model and the experimental
unitaries, and shot-to-shot errors due to fluctuations in the
experimental unitary.  While calibration errors contribute to the
TVD of the sampled distribution to the desired
boson sampling distribution, they can in principle be accounted for
by performing better calibration.  Shot-to-shot errors cannot be
overcome in this way and instead effectively result in sampling from a state
that is mixed, which can make the associated sampling task easier to
accomplish classically~\cite{oh_classical_2023}.

As discussed in the next section, and in SM Sec.~\ref{sup:calibration}, we expect that the
calibration errors in our system are larger than those in
state-of-the-art photonics experiments.  The precision in our
calibration is fundamentally limited by the number of experimental
trials that we can take before drifts in the Hamiltonian parameters
change the unitary applied (although we do not currently saturate
this limit).  The limitation in the precision of our calibration can be addressed by
improving the stability of the experiment, or by improving methods of
inference of the unitary.  We took a
first step in the direction of improving methods of inference by
introducing the maximum likelihood method described in the Methods
section ``Characterizing the single particle unitary.''

As discussed in SM Sec.~\ref{sup:error_model}, we expect that
shot-to-shot errors in our experiment are dominated by fluctuations
in the lattice depth.  We do not expect these errors to
significantly limit our experiment: when ignoring other sources of
errors, we estimate that shot-to-shot errors lead to a lower bound
on the fidelity of $F\gtrsim0.3$ after the quantum walk dynamics for
the 180 atom measurements performed in this work (see SM
Sec.~\ref{sup:error_model}). 

We estimate the combined contribution of shot-to-shot and 
calibration errors to the TVD to be below
$\sim0.01$ for the \(2\) and \(3\) atom measurements considered
in this section. These estimates take into account our model for
shot-to-shot errors (see SM Sec.~\ref{sup:error_model}), the uncertainty in the spectroscopic
calibration procedure (see SM Sec.~\ref{sup:calibration}), and the expected contribution to calibration
errors due to finite interaction strength (see SM
Sec.~\ref{sup:interactions}).

Our experimental distributions also differ from boson sampling
with number resolving readout because of parity projection.
We include parity projection in our models and do not consider this
to be a true discrepancy between the experimental probability distribution
and the models. However, we determined that the TVD between partity-projected
frequencies and number-resolved ones is \(\sim 0.05\) for both measurements appearing in this section.

\subsection{Comparing atomic and photonic implementations of boson sampling}

We compare some relevant figures of merit of state-of-the-art implementations of (different variants of) boson sampling in Tab.~\ref{etab:compare}. 
In photonics experiments, it can be challenging to generate and interfere large Fock states of photons due to transmission losses, and the probabilistic techniques that are often~\cite{peruzzo_quantum_2010, owens_two-photon_2011, sansoni_two-particle_2012, broome_photonic_2013, spring_boson_2013, tillmann_experimental_2013, bentivegna_experimental_2015, carolan_universal_2015} (but not always~\cite{loredo_boson_2017, wang_boson_2019}) used for single photon generation.
To circumvent this difficulty, recent experiments have performed modified versions of boson sampling that take advantage of more easily accessible non-classical states of light~\cite{lund_boson_2014, bentivegna_experimental_2015, zhong_quantum_2020, madsen_quantum_2022, deng_gaussian_2023} at the cost of requiring additional assumptions to support the claim that classical simulation of the resulting sampling problem is computationally hard~\cite{hamilton_gaussian_2017, quesada_gaussian_2018}. Additionally, recent theoretical results suggest that the presence of significant amounts of loss allow one to accomplish the implemented sampling tasks classically~\cite{oh_tensor_2023}.

The low loss, and high state preparation and detection fidelities, presented in this work enable studies of boson sampling that require fewer assumptions for the hardness of classical simulation~\cite{aaronson_computational_2011}.
In addition, we do not expect that the shot-to-shot errors in our experiment exceed those in photonics experiments due to optical path length fluctuations (either in the interferometer, or, in the case of Gaussian boson sampling, between the photon source and the interferometer)~\cite{zhong_quantum_2020,madsen_quantum_2022}.
However, it is more challenging to characterize the applied unitary $U$ to high precision in our experiment than in photonics experiments, likely leading to larger calibration errors.
In photonics, characterizations of $U$ can be performed efficiently and with very low noise by taking advantage of bright coherent states containing a macroscopic number of photons~\cite{zhong_quantum_2020, madsen_quantum_2022}.
We are not able to prepare equivalent states of atoms, and instead rely on characterizations based on few-particle quantum walks, which have significant experimental overhead, or on the indirect characterizations described in SM Sec.~\ref{sup:calibration}.

\subsection{Units and confidence intervals}

Unless otherwise noted, all error bars and uncertainties in this article and its supplementary information are provided as $\pm1\sigma$ confidence intervals.

\section{Acknowledgements}

We acknowledge fruitful discussions with Scott Aaronson, Nelson Darkwah Oppong, Ivan H. Deutsch, Jose P. D'Incao, Alexey V. Gorshkov, and Klaus Mølmer regarding the original manuscript, and with Nicolas Cerf, Leonardo Novo, and Beno\^{i}t Seron regarding recent advancements in generalized bunching.
We further thank Nelson Darkwah Oppong, Nicholas E. Frattini, Lee R. Liu, Klaus Mølmer, and Shuo Sun for close readings of the manuscript.
This work includes contributions of the National Institute of Standards and Technology, which are not subject to U.S. copyright. 
The use of trade, product and software names is for informational purposes only and does not imply endorsement or recommendation by the U.S. government.
\textbf{Funding:} This work was supported by the AFOSR (FA95501910079), ARO (W911NF1910223), the National Science Foundation Physics Frontier Center at JILA (1734006), and NIST.
S. Geller acknowledges support from the Professional Research Experience Program (PREP) operated jointly by NIST and the University of Colorado.
\textbf{Author contributions:} A.W.Y., W.J.E., N.S., and A.M.K. contributed to developing the experiments. A.W.Y. and S. Geller performed the analysis of the results. All authors contributed to interpreting the results and preparing the manuscript. \textbf{Competing interests:} Scott Glancy works as a consultant for Xanadu Quantum Technologies. All other authors declare no competing interests.
\textbf{Additional information:} Supplementary information is available for this paper.
Correspondence and requests for materials should be addressed to A.M.K.
Reprints and permissions information is available at www.nature.com/reprints.
\textbf{Data and materials availability:} Experimental data and code used in this work are available on Zenodo~\doi{10.5281/zenodo.10453016}.

\setcounter{section}{0}
\setcounter{equation}{0}
\setcounter{figure}{0}
\setcounter{table}{0}

\renewcommand\thetable{E\arabic{table}}
\renewcommand\theHtable{E\arabic{table}}
\setcounter{table}{0}

\renewcommand\thefigure{E\arabic{figure}}
\renewcommand\theHfigure{E\arabic{figure}}
\setcounter{figure}{0}

\renewcommand\theequation{E\arabic{section}.\arabic{equation}}
\renewcommand\theHequation{E\arabic{section}.\arabic{equation}}
\setcounter{equation}{0}


\begin{figure*}
	\includegraphics[width=\linewidth]{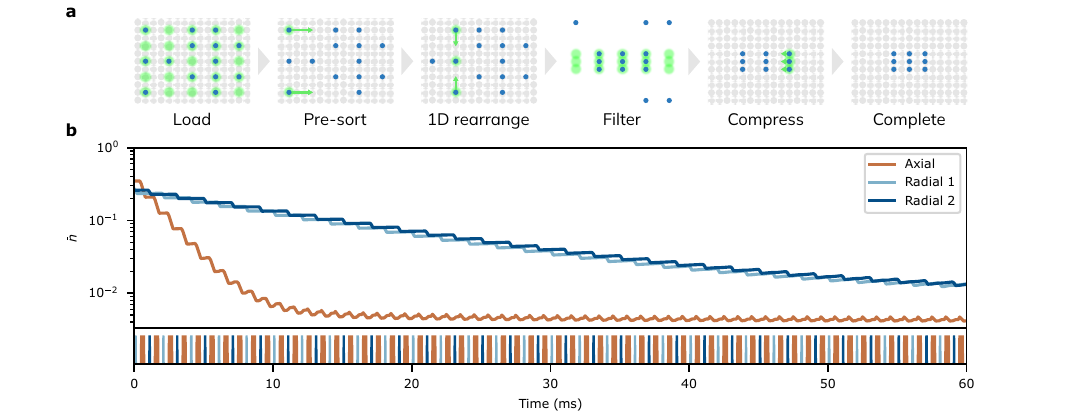}
	\caption{\textbf{State preparation.} \textbf{a}, Rearrangement of atoms (blue circles) in an optical lattice (grey circles denote sites in the lattice) using optical tweezers (green) must balance several conflicting requirements, leading to the multistep algorithm described in the Methods. \textbf{b}, To optically cool lattice-trapped atoms with high fidelity, we use a pulsed cooling sequence involving 0.4~ms axial cooling pulses, and 0.2~ms radial cooling pulses (timing diagram pictured in lower panel). We compute the expected average thermal occupation $\bar{n}$ as a function of time in each of three nearly-orthogonal axes of a given site in the lattice via a master equation calculation, yielding reasonable agreement with measured values in the experiment. Note that we optimize for high-fidelity cooling of the axial direction at the cost of slightly worse cooling in the radial directions.}
	\label{efig:1}
\end{figure*}

\begin{figure*}
	\includegraphics[width=\linewidth]{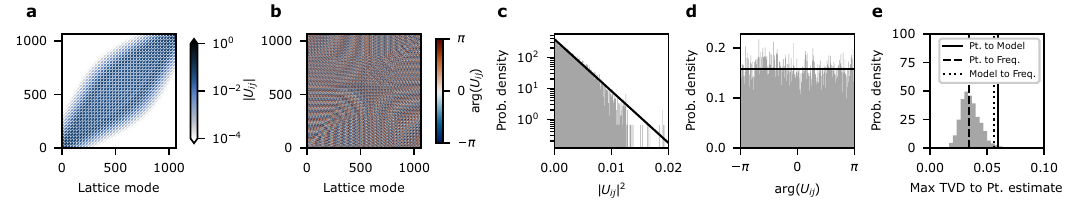}
	\caption{\textbf{Properties of the single particle unitary.} \textbf{a}, \textbf{b}, The single particle unitary $U$ is depicted here for an evolution time of $t = 6.45$~ms, as is relevant to the measurements in Fig.~\ref{fig:4}. The finite waist of the optical lattice beams and resulting harmonic confinement means that $U$ does not significantly couple all sites to each other, and thus is not Haar random.
  \textbf{c}, \textbf{d}, $U$ exhibits features of a Haar random matrix when considering only a $15\times15$-site region near the center of the lattice.
  In this region, the distribution of the norm-square of the amplitudes in $U$ are well-captured by the Porter-Thomas distribution for 385 outputs (black line in \textbf{c}).
  The distribution of the phases in $U$ is well-captured by the uniform distribution (black line in \textbf{d}).
  \textbf{e}, We perform maximum likelihood (ML) inference of a submatrix of the single-particle unitary based on one- and two-particle data (see Fig.~\ref{fig:3}), and compare the point estimate to ML estimates of bootstrap resamples of the data, and to the spectroscopic calibration (see SM Sec.~\ref{sup:calibration}).
  To quantify this comparison, we compute the one- and two-particle distributions generated by the inferred parameters, and compute the total variation distances (TVDs) of these distributions, then take the maximum of the TVDs over the prepared input patterns.
  We call this quantity the max TVD between two sets of distributions.
  The depicted histogram is the max TVD between the point estimate and the ML estimates of the bootstrap resampled data.
  Shown also are the max TVD between the frequencies of the data (Freq.) and the point estimate (Pt.), and that between the point estimate and the spectroscopic model (Model).
  The bootstrap histogram gives a sense of the size of the statistical fluctuation of the max TVD between the point estimate and the truth.
  The max TVD between the spectroscopic model and the point estimate is large compared to the bulk of the histogram, which is the expected behavior because statistical fluctuations in the model add to the statistical fluctuations in the point estimate.
  }
	\label{efig:2}
\end{figure*}

\begin{figure*}
	\includegraphics[width=\linewidth]{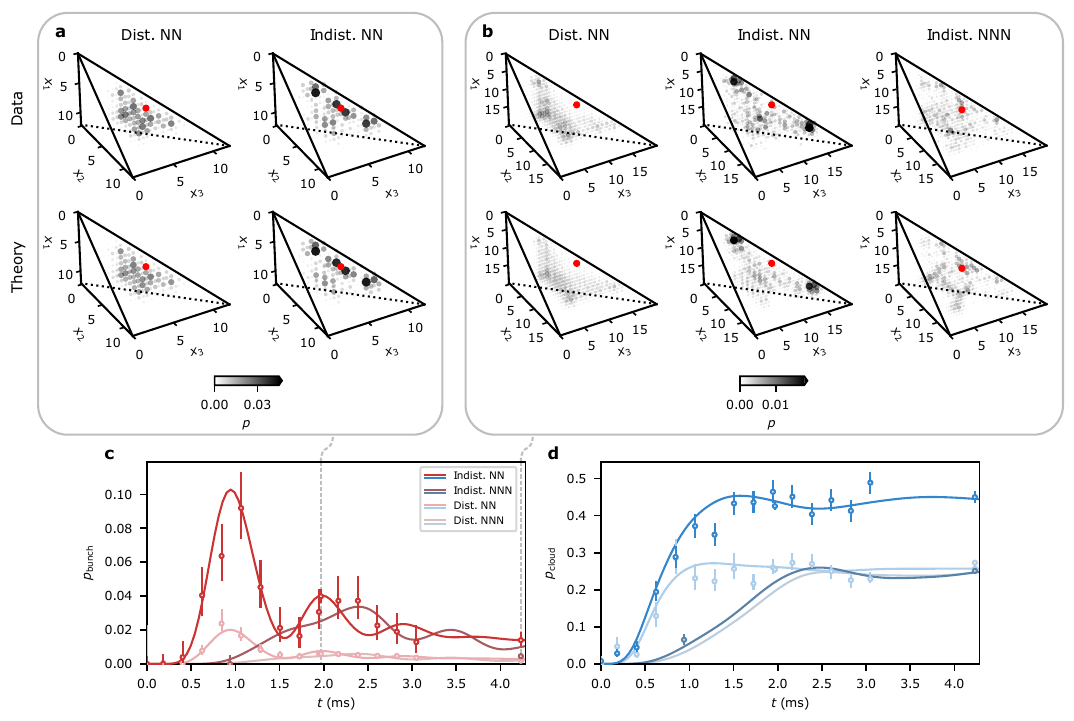}
	\caption{\textbf{Three particle quantum walks in 1D.} The output distributions resulting from three particle quantum walks at evolution times of \textbf{a}, 1.97~ms and \textbf{b}, 4.23~ms are in good agreement with theory.
  Similar to the two particle case, each three particle output can be uniquely labeled by the coordinates of the three particles $(x_1,x_2,x_3)$, with $x_3\le x_2\le x_1$.
  The probability $p$ of measuring an output state $(x_1,x_2,x_3)$ is indicated by both the size and color of the circle at the corresponding coordinates.
  The prepared input states are marked by the red disks, and include patterns with nearest-neighbor (NN) and next-nearest-neighbor (NNN) spacing.
  For NN input patterns, indistinguishable bosons (Indist.) exhibit enhanced probability to lie near the leading edge of the distribution along the main diagonal $(x_1=x_2=x_3)$ in comparison to distinguishable particles (Dist.).
  This tendency disappears for NNN input patterns.
  \textbf{c}, \textbf{d}, Like in the two particle case, we can coarse-grain the three particle distributions by measuring bunching and clouding, and find good agreement with theory as a function of evolution time.   
  All theory predictions in this figure correspond to error-free preparations of atoms with the appropriate particle statistics.
  }
	\label{efig:3}
\end{figure*}

\begin{figure*}
	\includegraphics[width=\linewidth]{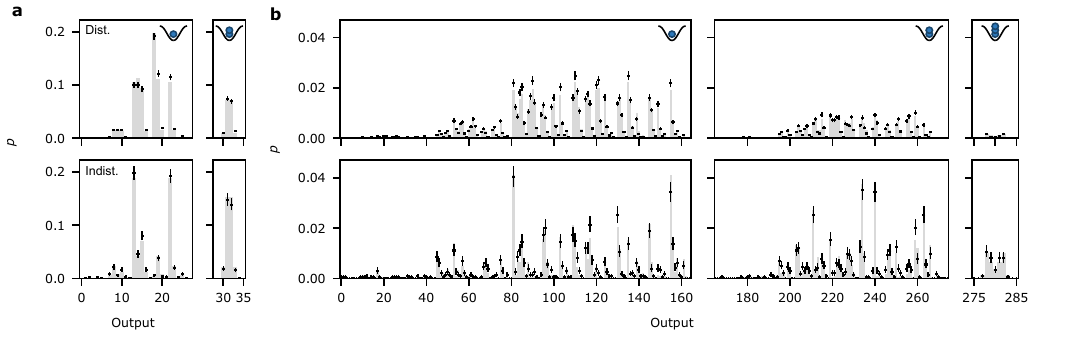}
	\caption{\textbf{Validation of two and three particle quantum walks in 1D.} The full output distribution after binning for \textbf{a}, two and \textbf{b}, three particles initialized at nearest-neighbor spacing, at an evolution time of $t_{\mathrm{HOM}}$ and $2t_{\mathrm{HOM}}$ respectively. The grey bars are theory for error-free state preparation, evolution, and detection with parity projection, and the black points are data. The upper row corresponds to distinguishable (Dist.) atoms, and the bottom row to unlabeled, nominally indistinguishable (Indist.), atoms. The outputs are grouped by the number of collisions (1, 2, or 3 atoms on the same site) that occur after binning, indicated by the inset cartoons. 
  }
	\label{efig:4}
\end{figure*}

\begin{table*}
  \centering
  \begin{tabular}{ccccccccccc} 
   \toprule
                                            & $n$ & $\mathcal{P}$ & $\mathcal{J}$ & $r$ & $m$ & Loss & \multicolumn{2}{c}{Detection} & Input & Evolution \\ 
   \midrule
  \cite{wang_boson_2019}                   & 20 & 0.975 &  0.93-0.954(1) & - & 60  & $26\;\%$ & 60-$82\;\%$ & Click & Fixed & Fixed \\
  \cite{zhong_quantum_2020}                & $50^*$ & 0.938 & - & 1.34-1.84 & 100 & $\sim45\;\%$ & 73-$92\;\%^\ddag$ & Click& Fixed & Fixed \\
  \cite{madsen_quantum_2022}               & $216^*$ & - & - & $\sim1.1$ & 216 & $\sim67\;\%$ & $95\;\%$ & Counting & Tunable & Tunable \\
  \cite{deng_gaussian_2023}                & $50^*$ & 0.962 & - & 1.2-1.6 & 144 & $57\;\%^\S$ & - & PPNRD & Phase & Fixed \\
  This work                                & 180 & $^\dag$ & $0.995^{+5}_{-16}$ & - & $\sim1015$ & 5.0(2)\% & $99.8(1)\;\%^\parallel$ & Parity & Pattern & Hamiltonian \\
   \bottomrule
  \end{tabular}
  \caption{
    \textbf{Comparison of large-scale boson sampling demonstrations.} For works involving Fock state boson sampling, $n$ denotes both the particle number and the number of input modes.
    Works involving Gaussian boson sampling are marked with a $^*$, in which case $n$ corresponds only to the number of input modes.
    $\mathcal{P} = 1-g^{(2)}(0)$ is typically referred to as the ``purity'' in photonics experiments, and is measured via second order correlations in Hanbury-Brown-Twiss-like experiments.
    $^\dag$To the extent that these measurements characterize the single-particle nature of the input field~\cite{wang_boson_2019}, in our experiments $\mathcal{P}\simeq1$ and is lower-bounded by our imaging fidelity of 0.998(1).
    However, our state purity is primarily limited by thermal motional excitations normal to the lattice, and can be estimated using the measured particle indistinguishability of $\mathcal{J} = 0.995^{+5}_{-16}$, which is an estimate of the purity assuming that the single-particle wavefunctions in the out-of-plane motional DOF are equal. 
    To characterize state preparation, we list $\mathcal{J}$ for Fock state boson sampling results and the squeezing parameter $r$ for Gaussian boson sampling results. 
    $m$ denotes the number of output modes in the linear optical network.
    ``Loss'' denotes the fraction of particles lost during evolution, including incoupling from the source to the linear-optical network, loss in the network, and outcoupling to the detectors. 
    Detection is characterized by the detection efficiency, and the type of measurement performed on each output mode. 
    ``Click'' refers to detecting the presence or absence of particles, ``parity'' to detecting particle number parity, ``PPNRD'' to pseudo-photon-number-resolving detection, and ``counting'' to full particle number-resolved readout.
    The work marked with $^\ddag$ includes fiber coupling loss in the estimate of detection efficiency, and the work marked with $^\S$ includes detection efficiency in the quoted value for loss.
    $^\parallel$The listed value for our work is a detection fidelity rather than an efficiency, and includes contributions from both particle loss and infidelity.
    Converting the other listed values to detection fidelities would involve including the effects of leakage light and dark counts, resulting in slightly lower values.
    ``Input'' refers to the class of states that can be prepared as inputs to the linear optical network, with ``phase'' referring to tunability of the phases of the prepared squeezed states, and ``pattern'' to nearly arbitrary Fock states with occupations of 0 or 1 on each input mode~(see Methods).
    ``Evolution'' refers to the family of linear optical networks that can be applied in a given system, with ``Hamiltonian'' referring to unitary evolution for variable time under a fixed Hamiltonian.
    For both ``input'' and ``evolution'', ``fixed'' refers to a single instance, and ``tunable'' to flexible, but not universal, programmability.
    The numbers appearing in this table are representative values for approximate comparison only, please refer to the original publications for details.
    } \label{etab:compare}
\end{table*}

\clearpage


\setcounter{section}{0}
\setcounter{equation}{0}
\setcounter{figure}{0}
\setcounter{table}{0}

\renewcommand\thetable{S\arabic{table}}
\renewcommand\theHtable{S\arabic{table}}
\setcounter{table}{0}

\renewcommand\thefigure{S\arabic{figure}}
\renewcommand\theHfigure{S\arabic{figure}}
\setcounter{figure}{0}

\renewcommand\theequation{S\arabic{section}.\arabic{equation}}
\renewcommand\theHequation{S\arabic{section}.\arabic{equation}}
\setcounter{equation}{0}

\renewcommand{\tocname}{Supplemental Materials}

\clearpage

\tableofcontents
\setcounter{secnumdepth}{2}

\incltocpage


\section{Theory of measuring indistinguishability with calibrated loss}
\label{sup:loss2indist}
In the Methods, we have described a method of measuring the indistinguishability of the atoms using a HOM measurement.
Here we describe a separate method of inferring the indistinguishability that takes different assumptions.
In particular, we assume here that the single particle losses are independent of whether one or two particles are prepared.
In the experiment, this is motivated by the physical model for this loss, which is primarily due to uncorrelated thermal occupation of higher in-plane bands.

The one particle experiments have the distribution
\begin{align}
  p_1(i|k) &= \left( 1-p(\emptyset|k) \right)|U_{i, k}|^2\\
p_1(\emptyset | k) &= p(\emptyset|k)
  \label{eq:oneparticle}
\end{align}
where $p(\emptyset|k)$ is the single particle loss probability starting in position $k$, $p_1(i|k)$ is the probability that an atom starting in position $k$ arrives in position $i$, and $U$ is the single particle unitary.
The distribution for two partially distinguishable particles with indistinguishability $\mathcal{J}$ subject to single particle loss and parity projection is
\begin{align}
  p_2(ij|kl) &= \left( 1-p(\emptyset|k) \right)\times\qquad\qquad \text{if $i\neq j$}\\
  &\left( 1-p(\emptyset|l) \right)p^{\text{partial}}(ij|kl)\nonumber\\
  p_2(i | kl) &= (1-p(\emptyset|k))p(\emptyset|l)|U_{i, k}|^2\nonumber\\
  + &(1-p(\emptyset|l))p(\emptyset|k)|U_{i,
  l}|^2\\
  p_2(\emptyset|kl) &= p(\emptyset|l)p(\emptyset|k)\nonumber\\
  + &\left( 1-p(\emptyset|k) \right)\left( 1-p(\emptyset|l)
  \right)\sum_{i}p^{\text{partial}}(ii|kl)
  \label{eq:twoparticle}
\end{align}
Our goal is to infer the parameter $\mathcal{J}$. We can plug in the expression from Eq.~\ref{eq:partial} into Eq.~\ref{eq:twoparticle} to isolate $\mathcal{J}$ as
\begin{align}
  \mathcal{J} &= \frac{p_2(\emptyset|kl) - p(\emptyset|l)p(\emptyset|k)
    - \sum_{i}p_1(i|k)p_1(i|l)}{\sum_{i}p_1(i|k)p_1(i|l)}
  \label{eq:estimate}
\end{align}
Using the plug-in estimator of the above expression we measure a value of $\mathcal{J} = 0.91^{+9}_{-13}$ for data that had a value of $\mathcal{J} = 0.972^{+10}_{-12}$ based on the HOM procedure outlined in the main text.
Note that the significantly larger errors in the estimate based on loss is the result of taking the ratio of two small numbers, and necessitates more precise single-particle calibration data than is necessary for the estimate based on the HOM measurement.
We did not correct for the bias in the plug-in estimate, because the statistical uncertainty in this estimate dominates any bias.

\section{Error model for unitary}
\label{sup:error_model}

Our primary error model for the unitary is that there are fluctuations in the laser power of the optical lattice.
This can be modelled as a Hamiltonian whose energy scale fluctuates, so 
\begin{align}
  H(s) &=  s H_0
  \label{eq:factoroutenegy}
\end{align}
where $s$ is an unknown unitless real number, and $H_0$ is some Hamiltonian with units of energy.
Now suppose further that the distribution $p(s)$ is known to be Gaussian,
\begin{align}
  p(s) &= \frac{1}{\sqrt{2\pi \sigma_s^2}}\exp\left( -\frac{(s - s_0)^2}{2 \sigma_s^2} \right)
  \label{eq:distn}
\end{align}
with known mean $s_0$ and known standard deviation $\sigma_s$.
Define $U(s) = e^{-i H(s) t}$, and suppose that $\ket{\psi}\!\!\bra{\psi}$ is some pure initial state on the same Hilbert space that $U(s)$ acts on.
We are interested in the fidelity of the resulting state to a pure target state $\Ket{\psi_0} = U(s_0)\ket{\psi}$ from acting by the ensemble of unitaries $U(s)$.
To that end, we first compute the state that results from applying the ensemble of unitaries.

The action of the ensemble gives
\begin{align}
  \rho' = \int_{-\infty}^\infty ds~p(s) U(s) \ket{\psi}\!\!\bra{\psi}U(s)^\dagger
  \label{eq:unitaryaction}
\end{align}
Now, make the transformation $s' = (s-s_0)/\sigma_s$, so that
\begin{align}
  \rho' = \int_{-\infty}^\infty ds'~\mathcal{N}(s') U'(s') \rho_0 U'(s')^\dagger
  \label{eq:nondimensionalized}
\end{align}
where $\mathcal{N}(s') = \exp(-s'^2/2)/\sqrt{2\pi}$ is the standard normal distribution, $\rho_0 = \ket{\psi_0}\!\!\bra{\psi_0}$ is the target state, and $U'(s') = U(\sigma_s s')$.

To perform the integral, first write $\rho_0$ in the eigenbasis of $H$ as $\rho_0 = \sum_{ik}\rho_{ik}\ket{i}\!\!\bra{k}$, and compute
\begin{align}
U'(s') \rho_0 U'(s')^\dagger
   &= \sum_{ik}\rho_{ik} e^{-i s'\sigma_s (\omega_i - \omega_k)t}\ket{i}\!\!\bra{k}
  \label{eq:klausderivation1}
\end{align}
where $\omega_i$ are the eigenvalues of $H_0$, so $H_0\Ket{i} = \omega_i \Ket{i}$.
Now since
\begin{align}
  &\int_{-\infty}^\infty e^{-i s'\sigma_s (\omega_i - \omega_k)t}\frac{e^{-\frac{s'^2}{2}}}{\sqrt{2\pi}} ds' \nonumber\\
  &= \exp\left( -\frac{(\sigma_s (\omega_i - \omega_k)t)^2}{2} \right),
  \label{eq:klausderivation2}
\end{align}
we can return to Eq.~\ref{eq:nondimensionalized} to obtain
\begin{align}
  \rho' =\sum_{ik}\rho_{ik}\exp\left( -\frac{(\sigma_s (\omega_i - \omega_k)t)^2}{2} \right)\ket{i}\!\!\bra{k}.
  \label{eq:klausderivation3}
\end{align}
The above display shows that the off-diagonal terms of the density matrix are damped by a Gaussian factor, where the coherences between far apart energies are damped more.

The fidelity to the target state is then
\begin{align}
  F &= \bra{\psi_0}\rho' \ket{\psi_0}\\
  &=\sum_{ii'kk'}\rho_{ik}\rho_{i'k'}\exp\left( -\frac{(\sigma_s (\omega_i - \omega_k)t)^2}{2} \right)\Tr(\ket{i}\!\!\bra{k} \ket{i'}\!\!\bra{k'})\\
  &=\sum_{kk'}|\rho_{k'k}|^2\exp\left( -\frac{(\sigma_s (\omega_{k'} - \omega_k)t)^2}{2} \right)
  \label{eq:fidelity}
\end{align}

Now, define $(\omega_{\text{max}} - \omega_{\text{min}}) = W$, where $\omega_{\text{max}}$ and $\omega_{\text{min}}$ are the maximum and minimum eigenvalues of $H_0$, respectively.
For example, if the Hamiltonian is a quantum walk Hamiltonian for a single particle on a line, where the spectrum is $\omega_k = 2J(\cos(k) - 1)$, we have $W = 4J$.
The quantity $W$ is called the bandwidth of $H_0$.
Then, the exponential in Eq.~\ref{eq:fidelity} can be bounded below as
\begin{align}
  \exp\left( -\frac{(\sigma_s (\omega_i - \omega_k)t)^2}{2} \right) &\ge \exp\left( -\frac{(\sigma_s Wt)^2}{2} \right)
  \label{eq:uniformbound}
\end{align}
So continuing the calculation in Eq~\ref{eq:fidelity}, we have
\begin{align}
  F &\ge \exp\left( -\frac{(\sigma_s Wt)^2}{2} \right) \sum_{k'k}|\rho_{k'k}|^2\\
  &= \exp\left( -\frac{(\sigma_s Wt)^2}{2} \right) 
  \label{eq:fidelitylower}
\end{align}
where in the last line we recognize that $\sum_{ik}|\rho_{ik}|^2$ is the purity of $\rho_0$, which is assumed to be $1$.

Now suppose that the Hamiltonian is linear optical on many bosons, where the fluctuating energy scale is correlated across all the particles.
That is, we can write the eigenbasis for the many-particle Hamiltonian as a site occupation list, and its action on that list is
\begin{align}
  H_n(s) \Ket{g_1, \ldots, g_m} &= s\sum_{x}g_x\omega_x \Ket{g_1, \ldots, g_m}
\end{align}
where $m$ is the dimension of the single particle Hamiltonian, the total number of particles is $\sum_{x}g_x = n$, and $s$ is still drawn from the distribution $p(s)$.
We define the many particle unitary as $U_n(s) = e^{-i H_n(s) t}$.

Then, if our many body initial state is $\tilde{\rho}$, we can write the state $\rho_n = U_n \tilde{\rho} U_n^\dagger$ in the eigenbasis of $H_n$ as
\begin{align}
  \rho_n = \sum_{g, g'}\rho_{g, g'}\ket{g_1, \ldots, g_m}\!\!\bra{g_1', \ldots, g_m'}
  \label{eq:manybodystate}
\end{align}
Now we perform a similar calculation as in the single particle case, to obtain
\begin{align}
  \int_{-\infty}^\infty &ds~p(s) U_n(s) \rho_n U_n(s)^\dagger = \\
 &\sum_{gg'}\rho_{gg'}\exp\left( -\frac{(\sigma_s \sum_{x}((g_x - g_x')\omega_x)t)^2}{2} \right)\ket{g}\!\!\bra{g'}
  \label{eq:manyparticleintegral}
\end{align}
Then to lower bound the exponential, observe that
\begin{align}
  \left|\sum_{x}((g_x - g_x')\omega_x)\right| \le n(\omega_{\text{max}} - \omega_{\text{min}}) = Wn \coloneqq W_n
  \label{eq:manybodybandwidth}
\end{align}
That is, the many body bandwidth $W_n$ grows proportionately to the number of particles $n$.
Now we can apply a similar lower bound as Eq.~\ref{eq:uniformbound} to obtain a lower bound of
\begin{align}
 F_n &\ge \exp\left( -\frac{(n\sigma_s Wt)^2}{2} \right)
  \label{eq:manybodybound}
\end{align}
on the fidelity of the many-particle state to the pure target state.
To generalize this to mixed states, suppose we start in some state $\tau = \sum_i q_i \psi_i$, where the states $\psi_i$ are pure, and $q_i$ is a probability distribution.
Then define $\rho_i' = \int^\infty_{-\infty}ds U(s)\psi_i U^\dagger(s) p(s)$.
Now we can apply concavity of the square root fidelity to obtain
\begin{align}
  \sqrt{F\left(\tau, \sum_i q_i \rho_i'\right)}&\ge  \sum_i q_i\sqrt{F( \psi_i, \rho_i')} \\
  &\ge \exp\left( -\frac{(n\sigma_s Wt)^2}{4} \right)
  \label{eq:concavity}
\end{align}
So that
\begin{align}
 F\left(\tau, \sum_i q_i \rho_i'\right) \ge \exp\left( -\frac{(n\sigma_s Wt)^2}{2} \right)
  \label{eq:mixedstatebound}
\end{align}
For our system, we have $W\approx1~\mathrm{kHz}$, and $\sigma_s\approx10^{-3}$, so for $n = 2$ and $t = 1.46~\mathrm{ms}$, the loss of fidelity due to this effect is negligible, and we therefore ignore it in the inference of the indistinguishability based on two particle experiments.
For $n = 180$ and $t = 6.45~$ms, we estimate that $F \gtrsim 0.3$ due to this effect (ignoring additional errors due to imperfect state preparation, measurement, and loss).

Conversely, an upper bound on fidelity depends on the specific states involved, where states with more coherences between far apart energy levels are damaged more.
The states considered in this work have nonnegligible coherences between all energy levels, and therefore we expect a loss of fidelity that increases with particle number.

\section{Confidence intervals for full bunching and clouding of distinguishable particles}
\label{sup:distinguishableconfint_bunchcloud}
Full bunching is defined to be the event that all particles arrive on the same site.
As discussed in the main text, the probability of full bunching for perfectly indistinguishable bosons is $n!$ larger than that for perfectly distinguishable particles.
We would therefore like to measure this probability, but due to parity projection, we cannot measure more than one particle on the same site.
The approach described in the main text to circumvent this issue is to appeal to the separability of the two spatial degrees of freedom in the quantum walk dynamics, and measure the number of particles that arrive in the same column, effectively creating a number-resolving detector with respect to the 1D quantum walk dynamics.
Nevertheless, if the particles actually end up on the same site in the full 2D lattice, some or all of them are lost.
We therefore must account for this effect when comparing to the probability of full bunching for distinguishable particles.

In the indistinguishable particle experiments, we cannot tell whether two particles were lost due to a single particle effect, or if they ended up on the same site and were lost due to parity projection.
Therefore, by postselecting on observing all the particles present, we are measuring the probability
\begin{align}
  p^b(\alpha|\neg \lambda \wedge \neg \chi)
  \label{eq:bosonpostselect}
\end{align}
where $\alpha$ is the event that all particles ended on the same column, $\lambda$ is the event that some particle was lost due to a single particle loss event, $\chi$ is the event that two particles arrived on the same site, $\neg$ is the logical ``not'', $\wedge$ is the logical ``and'', and $p^b$ is the probability distribution over outcomes when we prepare nominally indistinguishable particles.
Since the probability $\chi$ is dependent on the indistinguishability of the bosons, simply conditioning on the same event for distinguishable particles does not yield a quantity that can be directly compared to the postselected full bunching probability for the nominally indistinguishable particles.
We therefore instead would like to measure
\begin{align}
  \frac{p^d(\alpha \wedge \neg \lambda \wedge \neg \chi)}{p^b(\neg \lambda \wedge \neg \chi)}
  \label{eq:correctlynormalizeddisfb}
\end{align}
where $p^d$ is the probability distribution over outcomes when we prepare distinguishable particles.
This quantity can be directly compared to Eq.~\ref{eq:bosonpostselect}.

We would like to estimate the ratio in Eq.~\ref{eq:correctlynormalizeddisfb}.
Suppose we have an unbiased estimator $\hat{p}^d_{\alpha, 0}$ of the numerator $p^d(\alpha \wedge \neg \lambda \wedge \neg \chi)$.
Then we can construct the plug-in estimator of the ratio Eq.~\ref{eq:correctlynormalizeddisfb}, but it is biased because it is a ratio of probabilities.
To correct for this bias, we use the delta method~\cite{Shao2003}.
Concretely, our estimator is then
\begin{align}
  \hat{p}_{\alpha}^d &= \hat{p}^d_{\alpha, 0} / f^b(\neg \lambda \wedge \neg \chi) -\nonumber\\
  &\frac{1}{n^b}\left((1- f^b(\neg \lambda \wedge \neg \chi))f^b(\neg \lambda \wedge \neg \chi)\right) \times\nonumber\\ &\hat{p}^d_{\alpha, 0}/ f^b(\neg \lambda \wedge \neg \chi)^3
  \label{eq:fbdeltaestimator}
\end{align}
where $f^b(E)$ is the frequency of the event $E$ in the nominally indistinguishable particle preparation, $n^b$ is the number of experimental trials with that preparation.
This estimator is an estimator of Eq.~\ref{eq:correctlynormalizeddisfb} with bias of order $O(1/n_b^2)$.
Such an estimator is called first-order unbiased.
For our estimate of the clouding of distinguishable particles, we apply the same analysis, but the event $\alpha$ is replaced by the event $\eta$ that all the particles end up on the same half of the array.

It remains to specify the estimators $\hat{p}^d_{\alpha, 0}$ and $\hat{p}^d_{\eta, 0}$.
The  distinguishable particle bunching probability is
\begin{align}
  p^d(\alpha \wedge \neg \lambda \wedge \neg \chi) &= \sum_{x}\sum_{\mathbf{y} \in L_n}p(x, y_1|i_1)\cdots p(x, y_n|i_n)
  \label{eq:fbnocollision}
\end{align}
where $L_n$ is the set of subsets of size $n$ of the set of rows, and $p(x, y|i)$ is the probability that a single particle starting in site $i$ arrives in the site $x, y$.
This is the probability that all $n$ distinguishable particles arrive in the same column, but not in the same site.
Our task is then to estimate the above quantity from single particle experiments.

Let $l(j|i)$ be the number of trials where the outcome $j$ was observed, given the single particle preparation $i$.
Let $N_i$ be the total number of experimental trials taken in the preparation $i$, and let $f(j|i) = l(j|i) / N_i$ be the corresponding frequency of the outcome $j$ given preparation $i$.
Since Eq.~\ref{eq:fbnocollision} is a multilinear polynomial of the single-particle probabilities, the plug-in estimator is unbiased, so we define
\begin{align}
  \hat{r}^d_{\alpha, 0} &= \sum_{x}\sum_{\mathbf{y} \in L_n}f(x, y_1|i_1)\cdots f(x, y_n|i_n)
  \label{eq:pluginfb}
\end{align}
However, the sum in Eq.~\ref{eq:pluginfb} is too large to compute directly.
We therefore adopt a Monte Carlo approach to evaluating this sum.

The procedure is to sample with replacement $N_{MC}$ times from each of the single particle data $l(x_1|i_1), \ldots, l(x_n|i_n)$, and compute the frequency $f_{MC}(\alpha|i_1, \ldots, i_n) = n_{MC}(\alpha|i_1, \ldots, i_n)/N_{MC}$, where $n_{MC}(\alpha|i_1, \ldots, i_n)$ is the number of times that the event $\alpha$ occurs in the simulated data.
Then $f_{MC}(\alpha|i_1, \ldots, i_n) \rightarrow \hat{r}^d_{\alpha, 0}$ in the $N_{MC} \rightarrow \infty$ limit.
The speed of this convergence is given by the Chernoff-Hoeffding inequality, from which we can compute a reasonable value of $N_{MC}$.
Specifically, to estimate $\hat{r}^d_{\alpha, 0}$ to multiplicative error $\epsilon$ with probability $1-\delta$, we can take
\begin{align}
  N_\alpha(\delta, \epsilon) &= 2\log(1/\delta) / D(\hat{r}^d_{\alpha, 0}(1+\epsilon)|| \hat{r}^d_{\alpha, 0})
  \label{eq:nmc_from_chernoff}
\end{align}
many samples.
Here $D(a||b)$ is the KL divergence from the Bernoulli distribution with success probability $a$ to that with success probability $b$.
However, $\hat{r}^d_{\alpha, 0}$ is not known before the data is taken, so we use a model to compute a reference point $q^d_{\alpha, 0}$, and use that in place of $\hat{r}^d_{\alpha, 0}$ in the above display.
Then we define our estimator $\hat{p}^d_{\alpha, 0}$ to be this Monte Carlo approximation to $\hat{r}^d_{\alpha, 0}$.
We used the above analysis to construct a point estimate, and applied the bias-corrected percentile method to obtain a bootstrap confidence interval using 1000 bootstrap resamples.

When we took $\epsilon = .1$ and $\delta = .05$, we found that the number $N_\alpha(\delta, \epsilon)$ was too large to be computationally feasible when the number of atoms was larger than 3, so we instead took $N_{MC} = \mathrm{min}(N_\alpha(.05, .1), 4\times 10^6)$.
Limiting the number of Monte Carlo samples in this way makes our estimate less precise, but when we simulated the experiment, we found that the resulting bootstrap confidence intervals were small enough, so we decided to suffer this loss in precision.
The estimator $\hat{p}^d_{\eta, 0}$ was constructed similarly, with the event $\alpha$ replaced by the event $\eta$.
For the clouding probability, we chose $N_{MC} = \mathrm{min}(N_\eta(.05, .03), 1\times 10^7)$, because we found that we were satisfied with the size of the resulting bootstrap confidence intervals in simulation.

At $t=0$ in the bunching and clouding measurements, we expected from simulation that both the clouding and bunching probabilities would be equal to zero, so the above analysis would not yield a reliable confidence interval, since the bootstrap histogram would only contain mass at the value zero.
For this purpose, we performed no resampling of the distinguishable particle data, and instead concatenated the single particle samples to obtain distinguishable particle samples.
The confidence interval was computed from a union bound that combined Clopper-Pearson confidence intervals of the numerator and denominator in Eq.~\ref{eq:correctlynormalizeddisfb}.
Specifically, to compute the confidence upper bound of the  bunching probability, we computed a Clopper-Pearson upper bound of the numerator with significance $\alpha - \beta_u$, and a Clopper-Pearson lower bound of the denominator with significance $\beta_u$, and took their ratio to obtain a confidence upper bound of the ratio at significance level $\alpha$.
Here $\alpha$ was taken to be $(1-.68)/2$, and we computed that $\beta_u = .004$ was a reasonable value from simulations.
To compute the confidence lower bound, we computed a Clopper-Pearson lower bound of the numerator with significance $\alpha - \beta_l$, and a Clopper-Pearson upper bound of the denominator with significance $\beta$ and took the ratio to obtain a confidence lower bound of significance $\alpha$.
We took $\beta_l = .15999$ for the confidence lower bound.
We applied the same strategy for both the clouding and bunching probabilities at time $t=0$, with the same values of $\beta_u$ and $\beta_l$.

\section{Confidence intervals for generalized bunching and survival histograms}
\label{sup:distinguishableconfint_genbunch}

The generalized bunching statistic is not directly the probability of an event, but rather it is a linear combination of such quantities.
It is given in the indistinguishable particle case by
\begin{align}
  p^b_{\text{gb}} = \frac{1}{ {m \choose k}}\sum_{g}p^b\left(g \right){m-\#(g) \choose k - \#(g)}
  \label{eq:modifiedgenbunch}
\end{align}
where the sum is over all possible mode occupation lists whose entries are at most 1, and $\#(g)$ is the number of nonzero entries of the mode occupation list $g$.
Here $m$ is the maximum number of modes that the particles can occupy, and $k = \lfloor m - m/n\rceil$.

The analogous generalized bunching quantity for distinguishable particles is a multilinear function of the single particle probabilities, so the plug-in estimator is unbiased.
However, the sum over mode occupations is too large to calculate, and we therefore resort to a Monte Carlo method to estimate $p^d_{\text{gb}}$.
Like in section~\ref{sup:distinguishableconfint_bunchcloud}, the method is to sample with replacement $N_{MC}$ times from each of the single particle data $l(x_1|i_1), \ldots, l(x_n|i_n)$, and generate artificial distinguishable particle data.
We define $\Xi$ to be the parity projection operator, that takes a mode occupation list, and returns another mode occupation list, with $\Xi(g)_i = g_i~\%~2$, where $\%$ is the remainder operator.
Then for each artificial distinguishable particle sample, we compute ${m-\#(\Xi(g)) \choose k - \#(\Xi(g))}/{m \choose k}$, and take the empirical mean of the result.
This empirical mean is our estimator $\hat{p}^d_{\text{gb}}$.
It remains to specify the number of Monte Carlo samples.
The generalized bunching statistic is bounded between 0 and 1, so we can use the Hoeffding inequality to get an estimate of the number of samples to take.
To estimate $\hat{p}^d_{\text{gb}}$ to multiplicative error $\epsilon$ with probability $1-\delta$,
\begin{align}
  N_{\text{gb}}(\delta, \epsilon) &= \log(2/\delta) /\left(2 (\hat{p}^d_{\text{gb}}\epsilon)^2\right)
  \label{eq:nmchoeffgb}
\end{align}
many samples are sufficient.
Again, since $\hat{p}^d_{\text{gb}}$ is not known in advance, we use a model to calculate a reference value $q^d_{\text{gb}}$ that we use in place of $\hat{p}^d_{\text{gb}}$ in the above display.
We chose $N_{MC} = N_{\text{gb}}(.05, .02)$ to compute the generalized bunching statistic.

For the partially distinguishable data appearing in Fig.~\ref{fig:4}bc, we performed no resampling. 
Instead we concatenated the data into artificial partially distinguishable particle data, then computed Gaussian confidence intervals for generalized bunching, and independent Clopper-Pearson confidence intervals for each bin in the histograms of atom survival.

\section{Simulations of generalized bunching and atom survival}
\label{sup:genbunchsim}

The generalized bunching statistic and atom survival fraction depend on the number of sites that are occupied after the atom evolution, and are therefore sensitive to any effect that can change the number of observed atoms, including parity projection as well as single particle loss and detection errors.

Two kinds of detection errors can occur (in addition to any contribution to single particle loss due to false negatives).
A false positive in the first image after rearrangement can lead to an overestimate of the atom loss, since an atom that is missing from the input pattern is subsequently interpreted as being lost in the second image.
A false positive in the second image can lead to an underestimate of the atom loss, since an empty site can be observed to contain an atom.
The rate of the former error is lower for denser input patterns of atoms, which has the effect of reducing the observed atom loss and the measured value of $\overline{p_{\kappa}'}$ for data with fewer time labels, leading to more conservative measurements.
The rate of the latter error is similar for input patterns with different numbers of atoms.
As a result, we handle the effect of these errors by measuring their values from single atom calibration data.

For each relevant input site, we perform a calibration measurement with a single atom prepared at the appropriate site, and postselect on ``successful'' rearrangement (namely that, in the first image, the target site was observed to be occupied and no extra atoms were observed).
The combined effects of single particle loss and false positives in the first image are estimated by measuring the probability that no atom was observed in the second image, conditioned on no more than one atom being observed in the second image.
The effect of a false positive in the second image is estimated by measuring the probability that two atoms are observed when only one was prepared.
Since these errors are small we ignore the effect of higher order error processes that contribute to these signals.

To incorporate the above calibrations into our simulations, we take advantage of the fact that the absence of an atom, whether due to loss or detection errors, is expected to be an effect that commutes with the linear optical dynamics~(see SM Sec.~\ref{sup:interactions})~\cite{brod_classical_2020}.
Given this observation, our simulations proceed by simulating samples of the experiment without the above errors, applying single atom loss (including due to detection errors), then parity projection, and finally detection errors that lead to extra sites that are observed to contain atoms.

\section{Calibration of spatial variation in the lattice}
\label{sup:calibration}
\begin{figure*}
	\includegraphics[width=.85\linewidth]{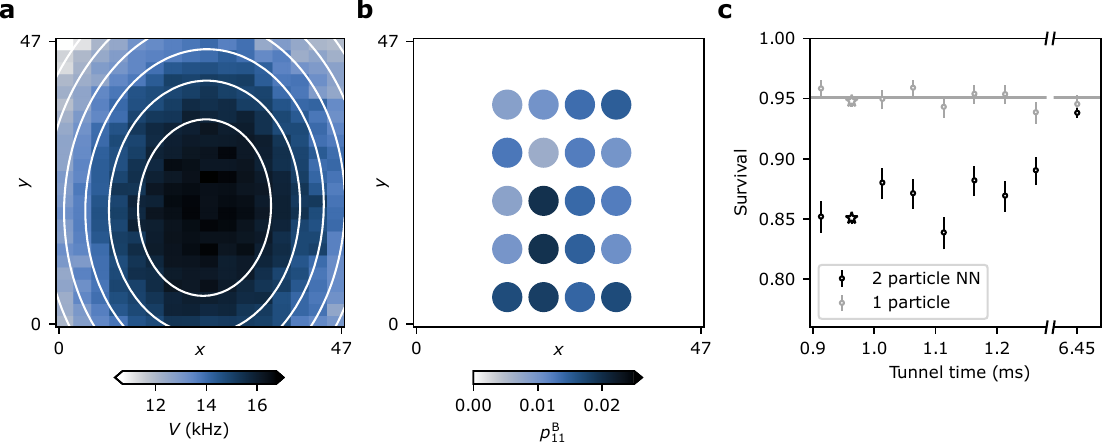}
	\caption{\textbf{Lattice calibration and loss.} \textbf{a}, Spectroscopic measurements of the lattice-induced AC Stark shift of the $\rm{{^1S_0} \leftrightarrow {^3P_1}}$ transition are used to characterize the lattice depth $V$ as a function of position in the lattice ($x$ and $y$ are in units of the lattice spacing), here pictured at the ratio of 2D and axial lattice depths used during tunneling. White contours correspond to a Gaussian fit to this potential, which is in good agreement with the expected lattice beam waists~\cite{young_tweezer-programmable_2022}. This fit is used in our model of the lattice Hamiltonian, and by extension our model of $U$. \textbf{b} HOM measurements at different locations in the lattice do not show a strong dependence on location over a region that spans all input sites used in this work (colored circles show the quality of the HOM dip at the corresponding position). Note that the HOM data appearing in Fig.~\ref{fig:1} and \ref{fig:2} of the main text is averaged over three regions centered at $x = 27$ and $y = (16.5, 24.5, 32.5)$. \textbf{c}, Single particle loss is not observed to vary over the evolution times in this work, here pictured at evolution times near the HOM dip, and at the latest evolution time explored in this work of 6.45~ms (horizontal line corresponds to the mean single particle loss). Data points marked by a star are taken at $t = t_{\mathrm{HOM}}$. The excess loss in the two particle data is associated with collisions between atoms during readout, which can be used as a probe of interference, as discussed in section~\ref{sup:loss2indist}. Notice that as the atoms spread out and the atom density goes down, the gap between the measured one and two particle loss closes.}
	\label{sfig:1}
\end{figure*}

To spectroscopically measure our lattice potential we change the angle of the magnetic field in the experiment such that the lattice is not magic for the $\rm{{^1S_0} \leftrightarrow {^3P_1}}$ transition~\cite{young_tweezer-programmable_2022}.
By measuring the local AC Stark shift across the lattice, we can measure the effective chemical potential on each site in the lattice, as shown in Fig.~\ref{sfig:1}a.
Note that this procedure is repeated separately for the axial and 2D lattices, as well as at the ratio of depths used during the tunneling experiments in the main text.
The change in lattice depth as a function of position affects not only the chemical potential, but also the tunneling energies.
To model this effect, we perform 2D band structure calculations to extract the tunneling energy as a function of lattice depth~\cite{young_tweezer-programmable_2022}, and combine these predictions with the measured chemical potential to construct a model of our Hamiltonian.
This relies on the approximation that the chemical potential varies smoothly across the lattice, which is reasonable given the results of the above spectroscopic measurements.

To confirm that the quality of our state preparation does not vary significantly across the lattice, we perform the HOM measurements described in the main text as a function of position in the lattice, and do not observe significant variation in performance across a region containing all input sites used in the main text (Fig.~\ref{sfig:1}b).
These HOM measurements can be supplemented by the loss measurements described in section~\ref{sup:loss2indist} (Fig.~\ref{sfig:1}c).

Note that we do not see increased single particle loss as a function of evolution time, even at the latest times explored in this work.
This suggests that the loss is exclusively due to filtering of imperfectly prepared atoms as described in the Methods, and does not scale with the depth of the effective linear optical network we apply.
It is important to note that avoiding the effects of such loss is critically important for maintaining the scaling of the computational hardness of the boson sampling problem~\cite{garcia-patron_simulating_2019}.
Ultimately, effects like parametric heating and vacuum lifetime will result in network depth-dependent single particle loss.
However, we expect these effects to become relevant only on the timescale of seconds~\cite{young_half-minute-scale_2020} in comparison to the millisecond-scale experiments explored in this work.

\section{The role of interactions}
\label{sup:interactions}

\begin{figure}
	\includegraphics[width=.85\linewidth]{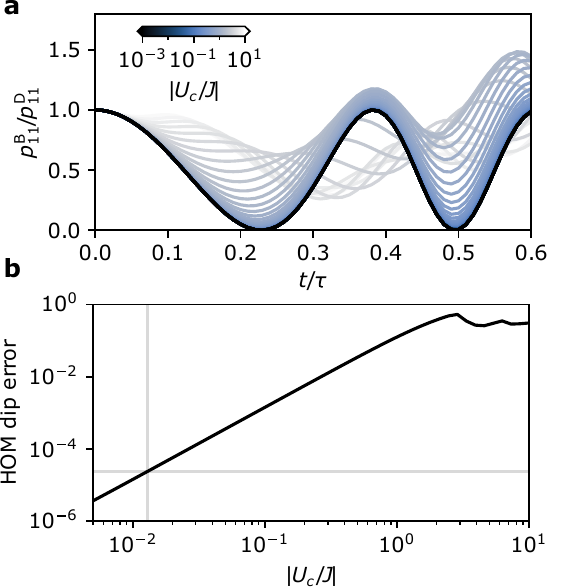}
	\caption{\textbf{The role of interactions.} \textbf{a}, We simulate the evolution of two nearest-neighbor atoms in a uniform 2D lattice under the influence of a variable contact interaction $U_c$ in comparison to the nearest-neighbor tunneling energy $J$. We define $\tau = 1/J$. \textbf{c}, The deviation in the visibility of the HOM dip from unity is characterized as a function of interaction strength, with the experimentally relevant values marked by the grey lines.}
	\label{sfig:2}
\end{figure}

The dominant elastic interactions in our experiment are due to $s$-wave scattering with a scattering length of $a = -1.4(6)a_0$~\cite{martinez_de_escobar_two-photon_2008}, where $a_0$ is the Bohr radius. Based on calculations of the band structure in our experiment~\cite{young_tweezer-programmable_2022} and the resulting Wannier functions, we expect this to result in contact interactions with a characteristic energy of $U_c/\hbar = -2\pi\times 1.7(7)$~Hz, which is about two orders of magnitude lower than $J$ in our experiments. Taking these interactions into account, our experiment is described by the Bose-Hubbard Hamiltonian:

\begin{align}
	H &= -\sum_{i,j} J_{ij} \hat{a}_i^\dagger \hat{a}_j - \sum_{i} V_i \hat{n}_i + \frac{U_c}{2}\sum_{i} \hat{n}_i (\hat{n}_i - 1)
	\label{eq:HamiltonianHubbard}
\end{align}

We simulate the resulting interacting dynamics exactly for two and three particles (Fig.~\ref{sfig:2}) via matrix exponentiation in \textsc{QuSpin}~\cite{weinberg_quspin_2019}, and compare these results to the non-interacting case. For all signals involving two and three particles, we find that the resulting errors are at or below the $10^{-4}$ level. 
Although we do not explicitly check if these interactions affect any of the signals at higher particle number, we do confirm that the measured signals are not significantly affected by triple occupation or higher of a given site. Since $n_i$ must reach a value of $\sim 10$ for the interaction term in Eq.~\ref{eq:HamiltonianHubbard} to be similar to the hopping energy for a single atom, we do not expect these interactions to significantly affect any of the signals at higher particle number~\cite{dufour_many-body_2020}.

Due to conservation of energy and momentum, the dominant inelastic interactions in our experiment involve three body processes. A loose upper bound for the three body loss rate is $\beta_{ggg} < 10^{-27}~\mathrm{cm^6 s^{-1}}$~\cite{sorrentino_laser_2006}, with an expected value closer to $\beta_{ggg} = 2.0(2)\times10^{-30}~\mathrm{cm^6 s^{-1}}$~\cite{goban_emergence_2018}. 
The above loss rates suggest that for three particles occupying a single site, the single-particle lifetime is several orders of magnitude longer than the evolution times explored in this work.
Additionally, for the particle densities at which we operate, it is rare for three particles to be simultaneously overlapped on a single site in the first place.
As a result, we conclude that the effect of inelastic collisions are negligible, and that all relevant loss processes that occur during the quantum walk dynamics can be described as a single-particle effect.

\section{Error budget}
\label{sup:budget}
\begin{table}
  \centering
  \begin{tabular}{cc} 
   \toprule
   Error source             & Simulated value ($\times 10^{-3}$) \\ 
   \midrule
   Imperfect beam splitter  & $24.1$ \\
   Axial cooling            & $7.3$ \\
   Averaging over regions   & $3$ \\
   Interactions             & $<0.1$ \\
   \midrule
   Total                    & $<0.035$ \\
   Measured                 & $0.029^{+15}_{-10}$ \\
   \bottomrule
  \end{tabular}
  \caption{\textbf{Error budget for particle indistinguishability.} For the HOM measurement appearing in the main text, the dominant error is due to imperfections in the applied beam splitter due to diagonal and next-nearest-neighbor tunneling terms in our lattice. The dominant error affecting state preparation is expected to be imperfect cooling of the out of plane, or axial, motional degree of freedom. Interactions and averaging over different inhomogeneous regions in the lattice are expected to contribute only marginally to imperfect characterization of the input state.}
  \label{tab:error}
\end{table}

The calculations and discussion in previous sections can be combined into an error budget for the quality of our state preparation, which appears in Table.~\ref{tab:error}. After correcting for the form of the unitary that we apply, the dominant source of error is imperfect cooling in the axial direction. Note that this is an error budget for the postselected HOM measurement we describe in the main text, which suppresses effects relating to single particle loss, imaging errors, and imperfect rearrangement. Single particle loss is measured to be $5.0(2)\;\%$, and the typical per-atom success probability for rearrangement is $\sim98\;\%$ when considering all experiments appearing in this work (but can be improved with optimization, as discussed in the Methods). This suggests that the on-demand success rate to perfectly prepare, evolve, and detect a single atom is approximately $92\;\%$. This number is primarily limited by the trade-offs made between cooling the axial and radial directions in this work, and could be improved further if optimizing for data-taking without postselection.

\section{Statistics}
\label{sup:stats}

Here, we summarize the statistics collected for the central data in this paper. For a complete accounting of the data, please visit the accompanying Zenodo repository~[to be included after publication].

Data at the HOM dip (Fig.\ref{fig:1}cd, Fig.\ref{fig:2}c-e) is averaged over three regions in the lattice. After postselection, there are 1022 trials of unlabeled data, 1087 trials of position-labeled data, and (1300, 1182) trials for the two starting locations in the time-labeled data.
For these measurements, we ensure that the overall position of the atom array agrees before and after rearrangement to minimize the effect of transient fluctuations in the alignment between the tweezer array and the lattice.
Not enforcing this condition adds $\sim1\;\%$ additional error to the HOM dip.
For the remaining distinguishable calibration data, we choose to use time labeling instead of position labeling due to more precise handling of collisions, and lower experimental overhead to reach the same precision.

Data for clouding and bunching as a function of atom number (Fig.\ref{fig:2}e) is averaged over (3, 2, 1, 1, 1) regions for measurements with (2, 3, 4, 5, 8) atoms.
For (2, 3, 4, 5, 8) atoms there are (1022, 2329, 1402, 1741, 2227) trials of unlabeled data, and ((1300, 1182), (892, 892, 892), (215, 220, 218, 218), (211, 217, 218, 204, 198), (73, 77, 80, 80, 80, 80, 75, 74)) trials of time-labeled data.
The inner brackets denote how much data was taken at each distinct starting position.
Note that the time-labeled calibration data for three particles consists of measurements only at one of the three starting positions, which is translated to generate calibrations for the remaining starting positions.
This was done for convenience, since this data is averaged over two separate regions in the lattice and thus variations on the scale of a single lattice site are not resolvable.
We perform separate measurements at each starting position for all other atom numbers, including the two atom measurement at the HOM dip.

To infer the unitary directly from quantum walk data (Fig.\ref{fig:3}, Extended Data Fig.\ref{efig:2}e) we collect (460, 449, 463, 457) trials for each of the four single particle measurements, and (877, 863, 871) trials for each of the three two-particle measurements. The measurements for unitary inference, and all subsequent measurements, are performed for a single region in the lattice.

For measurements of generalized bunching for intermediate particle numbers (Fig.\ref{fig:4}a) we collect (2494, 3170, 3031, 1367, 866) trials of unlabeled data for measurements with (4, 9, 16, 25, 36) atoms.
For (4, 9, 16, 25, 36) atoms there are between (846-939, 530-939, 645-788, 420-470, 420-470) trials for each starting position in the time-labeled calibration data.
For (16, 25, 36) atoms we collect additional calibration data at the central starting position, for a total of (1465, 1153, 1153) trials.
Note that the 9 atom calibration data contains the 4 atom calibration data as a subset because these measurements were performed close together.
The same is true of the 36 and 25 atom calibration data. Otherwise, all calibration data sets are gathered independent of each other to ensure that the time-labeled calibration measurements are taken in close proximity to the associated unlabeled measurements.

For the 180 atom measurements (Fig.\ref{fig:4}bc) with (1, 2, 4, 9) time labels, we collect ((3266), (795, 770), (426, 334, 250, 419), (368, 495, 351, 364, 353, 358, 386, 395, 429)) trials.
The inner brackets denote how much data was taken for each distinct input pattern.
To compute the value of generalized bunching for this data, we choose $m = 1015$.
Note that although the 180 atom measurements are not postselected for perfect atom rearrangement, we do enforce that $>90\;\%$ of the input sites are populated to exclude rare outlier events where the rearrangement performs poorly.

\end{document}